\documentclass{aa}  

\usepackage{graphicx}
\usepackage{txfonts}
\usepackage{url}
\usepackage{hyperref}
\usepackage[]{datetime}
\usepackage{rotating}
\usepackage{multirow}
\usepackage{diagbox}
\usepackage{xcolor}
\usepackage{amsmath,amssymb}
\usepackage{cases}

\providecommand{\changed}[1]{{#1}} 

\providecommand{\given}{\ensuremath{\hspace{0.05em}\mid\hspace{0.05em}}}
\providecommand{\snr}{S/N}
\providecommand{\snrs}{S/Ns}

\providecommand{\msun}{\ensuremath{{\cal M}_\odot}}
\providecommand{\mjup}{\ensuremath{{\cal M}_{\rm J}}}
\providecommand{\yr}{\ensuremath{\mathrm{yr}}}
\providecommand{\au}{\ensuremath{\mathrm{au}}}
\providecommand{\pc}{\ensuremath{\mathrm{pc}}}

\providecommand{\mass}[1]{\ensuremath{{\cal M}_{#1}}}
\providecommand{\met}{\ensuremath{\mathrm{[M/H]}}}
\providecommand{\age}{\ensuremath{\tau}}
\providecommand{\sma}[1]{\ensuremath{a_{#1}}}
\providecommand{\flux}[1]{\ensuremath{f_{#1}}}
\providecommand{\period}{\ensuremath{T}}

\providecommand{\parallax}{\ensuremath{\varpi}}

\providecommand{\mpers}{\ensuremath{\textrm{m\,s}^{-1}}}

\providecommand{\mas}{\ensuremath{\textrm{mas}}}
\providecommand{\msun}{\ensuremath{M_\odot}}

\providecommand{\frr}{\ensuremath{R}}

\providecommand{\gaia}{Gaia\xspace}
\providecommand{\gdr}[1]{Gaia~DR{#1}\xspace}

\providecommand{\absG}{\ensuremath{M_{\rm G}}\xspace}
\providecommand{\absB}{\ensuremath{M_{\rm B}}\xspace}
\providecommand{\absR}{\ensuremath{M_{\rm R}}\xspace}

\begin{document} 

\title{Component masses in stellar and substellar binaries from Gaia astrometry and photometry}
\titlerunning{Binary component masses from Gaia}
\authorrunning{Bailer-Jones \& Kreidberg}

\author{C.A.L.\ Bailer-Jones \& L.\ Kreidberg}
\institute{Max Planck Institute for Astronomy, Heidelberg, Germany}

\date{Submitted 2026-01-16; Revised 2026-02-20; Accepted 2026-03-03}

\abstract{
The masses of stars and planets can be measured dynamically in binary systems.
For an unresolved binary, time series astrometry yields some orbital parameters, but
it cannot provide the component masses, because we observe only the motion of the system's photocentre.
However, as a star's luminosity is related to its mass, the observable photometry of both components together provides information on the system mass.
Here we develop a method to determine the individual component masses of an unresolved binary using the astrometric orbit together with three-band photometry from \gaia. 
We use a mass--flux relation fitted from stellar isochrone models for each \gaia\ band to infer the unknown flux ratio.
This enables our method to identify near equal-mass stellar binaries, expected to be the dominant source of false positive exoplanet candidates, without the need for additional follow-up.
Using a likelihood approach, we sample the posterior probability distribution over the stellar parameters, marginalizing over system age and metallicity.
We apply this to 20\,000 systems with a main sequence primary within 300\,pc of the Sun using data from the \gaia\ data release 3 non-single star catalogue.
Primary masses can be determined with a precision (1$\sigma$ posterior width) of 10--20\% in 90\% of cases.
Secondary masses, which extend down to \changed{planetary-mass objects}, are less precise, although half are more than 25\% precise.
Interestingly, adding either infrared photometry or spectroscopic orbits from \gaia\ does not change the mass estimates much \changed{(less than 4\% and 1\% respectively)}.
Interstellar extinction likewise has little impact for this sample.
We provide a catalogue of our mass estimates.
This work shows that reasonably precise masses can be obtained for stars and substellar objects using just the \gaia\ astrometry and photometry.
}
\keywords{}

\maketitle
\nolinenumbers

\section{Introduction}\label{sec:introduction}

The mass of an astronomical object is one of its most fundamental properties.
Masses can be measured in a variety of ways, but these all rely ultimately on a calibration via dynamical measurements.
Observations of the mutual orbit of two gravitationally-bound bodies provides the means to determine the masses of both components.

The \gaia\ mission provides the data to determine orbits for millions of systems \citep{2016A&A...595A...1G}.
As part of its third data release \citep{2023A&A...674A...1G},
the \gaia\ Data Processing and Analysis Consortium (DPAC) has published solutions for about 840\,000 spatially-unresolved non-single stars (NSS).
Of these, around 443\,000 have two-body Keplerian orbit solutions.
This includes astrometric, spectroscopic, and eclipsing binaries as well as combinations of these. 

Orbital data alone only allow us to infer the masses of both components in a limited set of cases, namely when we have radial velocities of both components (double-lined spectroscopic binaries, SB2) together with either astrometric or eclipsing orbits \citep{2020svos.conf..329S}.
This is unfortunate, because SB2 data is only available for systems that are bright enough to have good single-epoch radial velocity measurements and where the secondary is not too faint (mass ratio not too large).
It would be helpful if we could obtain masses just from astrometric orbits, because these are available for many more systems, including fainter sources, sources in crowded fields, and highly inclined orbits. To achieve this we need additional information on the component masses. This we can obtain from broad band photometry.

For insight into how photometry helps infer mass, an unresolved binary comprising two identical stars is twice as bright as just one such star, and will therefore lie 0.75\,mag higher in the colour--absolute--magnitude diagram (CAMD). To identify this shift we need to convert the observed apparent magnitude into the absolute magnitude. We can do this via the parallax, which \gaia\ determines self-consistently along with the NSS astrometric orbit. For an unequal-mass binary, the shift in the CAMD will be different. But as there is a relationship between mass and absolute magnitude for each component of the binary, the magnitude shift in the CAMD contains useable information on the system's mass ratio.
The absolute magnitude does not only depend on the mass, however, so using (PARSEC) stellar models, we can fit a relationship between stellar mass, metallicity, and age on the one hand and absolute magnitude on the other. 

\begin{figure}[t]
\begin{center}
\includegraphics[width=0.40\textwidth, angle=0]{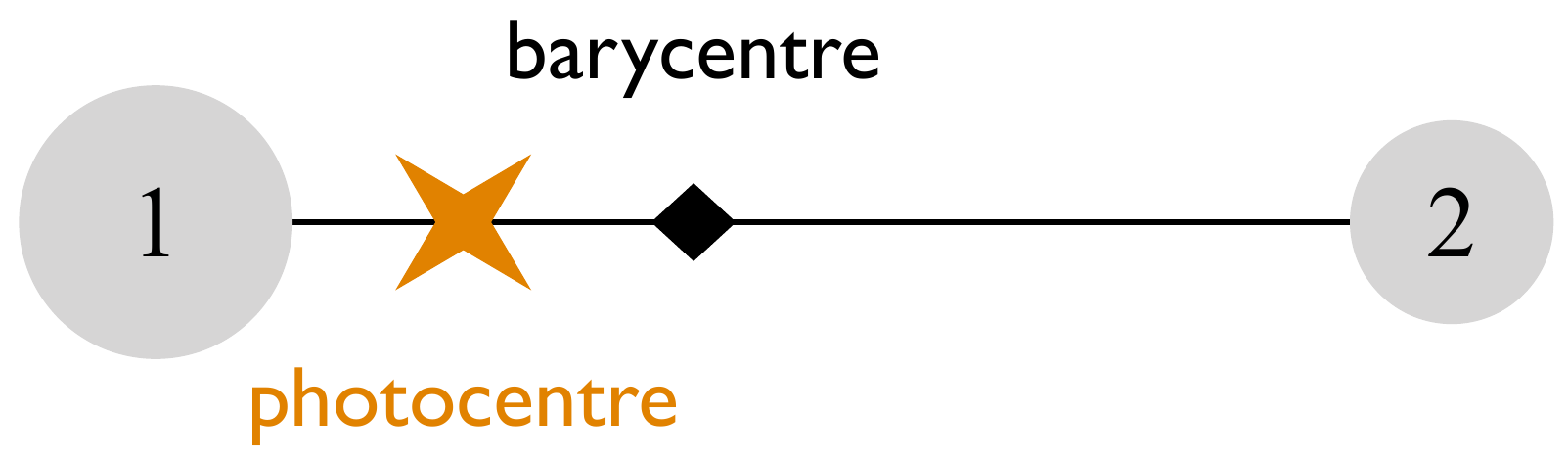}
\caption{Sketch of two objects, 1 and 2, of different masses in orbit about their barycentre, with a flux ratio that places the photocentre away from the barycentre. In an unresolved binary we observe the motion of the photocentre about the barycentre.
\label{fig:binary_sketch} }
\end{center}
\end{figure}

In this paper we use the NSS astrometric orbit together with the parallax and three-band \gaia\ photometry
(G, BP, RP)
to estimate the masses of both components of an unresolved binary.
For such a binary, \gaia\ can only observe the motion of the system's photocentre, so the measured 
semi-major axis (SMA) depends not only on the components' mass ratio but also on their flux ratio (see figure~\ref{fig:binary_sketch}).
We can predict this flux ratio for given stellar parameters from the stellar model, subject to the constraint that their summed flux agrees with the observed flux.
Assuming a common metallicity and age for the two components, this gives us four stellar parameters to infer from the three-band photometry (each for the two components together), the parallax, the photometric SMA, and the orbital period.
We perform this inference by defining the likelihood of the data given the stellar parameters, adopting a broad prior, and then sampling the resulting posterior probability density function (PDF).
The age and metallicity are only weakly determined by the photometry -- and not at all by the astrometry -- but their presence is necessary to account for the broad scatter in the mass--absolute magnitude relationship.

Although NSS refers to its subject as non-single stars, there is nothing in the \gaia\ measurement technique that precludes the detection of substellar mass companions, namely brown dwarfs or exoplanets. These are, however, harder to detect reliably on account of their low masses, and therefore a smaller perturbation of the more massive primary star.  When the companion is also very faint, inference is made easier by assuming the companion contributes no light to the observed photometry. But we cannot make this assumption a priori. This is relevant because a faint low-mass object around a bright massive object has an identical astrometric orbit as a stellar binary comprising two stars of similar mass and brightness. By neglecting the light from the companion we force the former solution. We should instead permit a variable flux ratio in the inference.

Several other studies have estimated masses, or limits thereon, of unresolved binaries using the \gaia\ astrometry.
As part of the \gdr3\ validation, \cite{2023A&A...674A..34G} explored estimating masses for one or both components of the system for several different types of NSS solution.
Similar in spirit to our work, \cite{2024A&A...682A..12P} adopted a mass--luminosity relation to help estimate masses. This exploited the binary star results provided by the Multiple Star Classifier (MSC) in the \gaia\ processing \citep{2023A&A...674A..28F}, which estimates stellar parameters of unresolved binaries using their combined BP/RP spectrum.
Other studies have combined \gaia\ data with astrometry from Hipparcos and/or radial velocities from other surveys, often to extend the temporal baseline \citep[e.g.][]{2025ApJS..280...61A}
The \gaia\ data have also been used to identify black hole \citep{2023MNRAS.518.1057E,2024A&A...686L...2G}
and brown dwarf \citep{2024A&A...688A..44W,2025MNRAS.536.2485W} 
companions to stars.

The paper is laid out as follows.
In section~\ref{sec:methods} we describe our inference method. We summarize the results
in section~\ref{sec:results_main} and compare these to other estimates in section~\ref{sec:comparison}. We explore the impact of adding infrared photometry and spectroscopic orbital information to the inference in sections~\ref{sec:results_withNIR_noSpectro} and~\ref{sec:results_noNIR_withSpectro} respectively. We finish up with a discussion of some of the limitations and caveats in section~\ref{sec:discussion} and conclude in section~\ref{sec:conclusions}. Appendix~\ref{sec:performance_simulated_binaries} presents the application of our method to simulated data to give a lower bound on the performance. Appendix~\ref{sec:contamination} investigates the statistical contamination of stellar companions when looking for exoplanet companions.

\section{Methods}\label{sec:methods}

\subsection{What astrometry tells us}

Kepler's third law relates the physical SMA \sma{} to the period \period\ and the sum of the masses of the two components:
\begin{alignat}{2}
  \left(\frac{\sma{}}{\au}\right)^3 \,&=\,
  \left(\frac{\period}{\yr}\right)^2
  \left(\frac{\mass1+\mass2}{\msun}\right)
\label{eqn:kepler3}
\end{alignat}
where \yr\ is approximately the Earth's sidereal year.
From the \gaia\ epoch astrometry one infers not the physical SMA but rather the
SMA of the photocentre, \sma{p}, which is related to \sma{}\ via \citep{1956VA......2.1040V}
\begin{alignat}{2}
\sma{p} \,&=\, \sma{} \, \left|\frac{1}{1 + \mass1/\mass2} - \frac{1}{1 + \flux1/\flux2} \right|
\label{eqn:sma_photocentre}
\end{alignat}
where \flux1\ and \flux2\ are the fluxes of the two components in the band in which the astrometry is measured (in our case the \gaia\ G band). Note that this expression in invariant under a permutation of the subscripts.

DPAC used the \gaia\ epoch astrometry to compute both $T$ and \sma{p}. The NSS catalogue actually lists the orbital parameters as Thiele--Innes parameters (and their covariances) instead of the more familiar Campbell parameters, but \cite{2023A&A...674A...9H} shows how to compute \sma{p}\ and its uncertainty from these (and provides a python tool to do it). They provide the Thiele--Innes parameters in angular units (\mas), so we divide by the parallax to put this in linear units (\au). This assumes the distance equals the inverse parallax which, although generally not true  \citep{2015PASP..127..994B}, is a good approximation for the vast majority of the systems we analyse due to their high parallax signal-to-noise ratio (\snr).

Combining equations~\ref{eqn:kepler3} and~\ref{eqn:sma_photocentre} gives
\begin{alignat}{2}
  \left(\frac{\sma{p}}{\au}\right)^3 \left(\frac{\period}{\yr}\right)^{-2} \,&=\,
  \left(\frac{\mass1+\mass2}{\msun}\right)
  \left|\frac{1}{1 + \mass1/\mass2} - \frac{1}{1 + \flux1/\flux2} \right|^3 \ .
\label{eqn:kepler3_photocentre}
\end{alignat}
The \gaia\ astrometry gives the left side of this equation, which we call the `astrometric measurement' and denote with symbol $A$.
When $\flux2 \ll \flux1$, the right side of this equation is independent of the fluxes, so we get an estimate of a function of \mass1\ and \mass2.
For this reason the right side is sometimes called the `astrometric mass function'. (If we additionally assumed $\mass2 \ll \mass1$, then the equation becomes $\mass2^3 \simeq A \mass1^2$.)
To get estimates of each mass individually we need at least one more mass-sensitive measurement, such as absolute magnitude in a passband or the radial velocity variation.

\subsection{Introducing photometry to break the mass degeneracy}\label{sec:breaking_degeneracy}

The simplification of neglecting the flux of the secondary is often made 
when looking for faint, low mass companions to stars, as this leads to a much simpler approach.
But this overlooks the degeneracy between mass and flux present in equation~\ref{eqn:kepler3_photocentre}.
\changed{Two stars of near-equal masses and near-equal fluxes can produce an arbitrarily small astrometric measurement.
Yet two objects with the right combination of large mass and flux ratios, such as a star and a planet, can produce the same astrometric measurement.}
So unless we can constrain the fluxes, a given astrometric measurement can correspond to any combination of masses.

Our model accommodates the general case where \flux2\ is not negligible. 
Clearly it would be useful to know the flux from each component, but because the system is unresolved we can only measure the sum of their fluxes.
If absolute magnitude depended only on stellar mass in some known way, then the sum of the fluxes together with the parallax (to convert flux to absolute magnitude), along with the astrometric measurement, would allow us to solve for the two masses (two equations, two unknowns).\footnote{This is only possible for a limited set of functional dependencies of flux on mass, and even then there can be multiple solutions.}
In reality, though, the absolute magnitude on the main sequence is also a function of age, metallicity, and other parameters, so there is not a one-to-one relationship between mass and absolute magnitude.
This means we need the absolute magnitude in more than one band in order to estimate individual masses (and other parameters).
To do this we construct a forward model of the absolute magnitude in each band as a function of mass, age, and metallicity.
These, together with equation~\ref{eqn:kepler3_photocentre}, form a set of nonlinear equations which can be solved numerically.

Real stars and noisy measurements don't conform precisely to a set of deterministic equations, however, so these won't give a solution. We therefore adopt the likelihood approach in which we define the probability of the measurements given the masses (and other parameters) and then seek the most likely parameters. Rather than simply maximizing the likelihood, we define and sample the posterior probability density function (PDF), which is the product of the likelihood with a prior over the parameters. By sampling the posterior we obtain a point estimate of the parameters and a confidence interval on that estimate.

\subsection{Photometric forward models from PARSEC}\label{sec:forward_model}

\begin{figure}[t]
\begin{center}
\includegraphics[width=0.49\textwidth, angle=0]{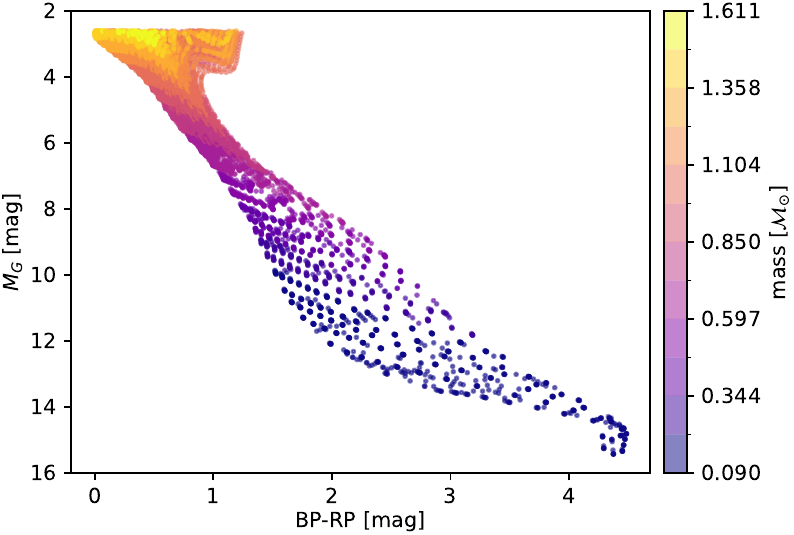}
\caption{Colour--absolute magnitude diagram of the PARSEC models used to fit the forward models of equation~\ref{eqn:forward_model}.
\label{fig:CQD_forward_model_data} }
\end{center}
\end{figure}

We use PARSEC1.2S stellar models (\citealt{2012MNRAS.427..127B}) to define the relationship between absolute magnitude ($M$) in each of the three \gaia\ bands $G$, $B$ (=BP), and $R$ (=RP) and the stellar parameters mass \mass{}, age \age, and metallicity \met.
\changed{PARSEC is a stellar evolution code that solves the equations of stellar structure using information on nuclear reaction rates, material opacities, and transport processes, to predict the variation of stellar properties as a function of time.
We used the public interface of PARSEC to generate a synthetic grid of stars} over the mass range 0.09--5.0\,\msun\ with a variable mass step size, a $\log_{10}(\age/\yr)$ range of 7.6--10.0 in steps of 0.15\,dex, and a metallicity range of $-2.0$ to 0.25\,dex in steps of 0.15\,dex, retaining only evolutionary stages 1 (main sequence), 2 (sub giant branch), and 3 (red giant branch).
To fit the forward models we select only those sources within the box $2.6 < \absG < 15.4 $ and $0.0 < \absB - \absR < 4.5$. This reduced the mass range to 0.09--1.61\,\msun\ (the age and metallicity ranges were unchanged) and left us with a 
set of 7985 models for fitting the forward models. These are shown in Figure~\ref{fig:CQD_forward_model_data}. The majority (7333) are main sequence stars.

The forward model for the absolute magnitude is defined for each band independently by the three-dimensional function
\begin{alignat}{2}
  M_{i,X} \,&=\, y_X(\mass{i}, \age, \met)
    \label{eqn:forward_model}
\end{alignat}
where $X$ is one of $G, B$ and $R$, and $i=1$ or $2$ to indicate the primary and secondary component respectively. We experimented with various tree and neural network architectures for fitting these models, concluding on gradient boosted trees as implemented by GradientBoostingRegressor in the scikit-learn python package \cite{scikit-learn}.
We used a squared error loss function with 200 boosting stages.
85\% of the PARSEC models were used for training, the rest for testing.
The median absolute deviation of the fits are 0.013\,mag, 0.014\,mag, and 0.010\,mag in $G$, $B$, and $R$ respectively. The bias of each fit is less than 0.001\,mag in all cases.

\begin{figure}[t]
\begin{center}
\includegraphics[width=0.49\textwidth, angle=0]{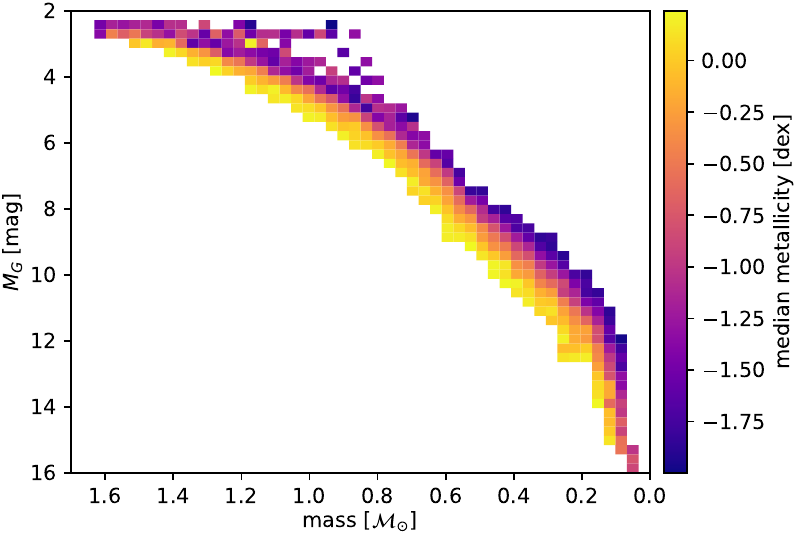}
\caption{Variation of $M_G$ with mass predicted by the fitted model of equation~\ref{eqn:forward_model}, colour-coded by median metallicity. 
\label{fig:forward_model_MG_vs_mass_met_colourcoded} }
\end{center}
\end{figure}

To illustrate the models we make 4000 stars by drawing log mass, log age, and metallicity from uniform distributions over the range of the fits and simulate photometry for each. The variation of $M_G$ with mass colour-coded by metallicity is shown in figure~\ref{fig:forward_model_MG_vs_mass_met_colourcoded}.

The forward models predict the absolute magnitudes for each component, from which we can compute the corresponding flux from
\begin{alignat}{2}
\hat{f}_{i,X} \,&=\, 10^{-0.4*(M_{i,X} + z_X - 5[1+\log_{10}\parallax])}
\label{eqn:apparent_flux_from_absmag}
\end{alignat}
where \parallax\ is the system parallax in arcseconds and $z_X$ is the magnitude zeropoint of the \gaia\ photometric system\footnote{Table 5.4 of the \gdr3\ online documentation from \url{https://gea.esac.esa.int/archive/documentation/GDR3/Data_processing/chap_cu5pho/cu5pho_sec_photProc/cu5pho_ssec_photCal.html}}. This assumes there is no interstellar extinction, which is a reasonable approximation for many stars within a couple of hundred of parsecs, but may be a source of error for more distant stars. We assess this assumption in section~\ref{results:extinction_correction}. Equation~\ref{eqn:apparent_flux_from_absmag} also assumes the parallax is the inverse of the distance which, 
because the parallax \snr\ is large for most systems in this study, is not a limiting approximation.

The lower mass limit of the grid is 0.09\,\msun, but we would like to be able to infer the masses of lower mass objects.
As such low-mass objects contribute little flux, it might be sufficient to simply adopt zero flux contribution when the proposed mass falls below this grid limit during the inference. But this is problematic when both components in a system are below this limit \changed{(although \gaia\ sees such systems only when they are very nearby)}. In lieu of reliable grids for lower mass objects, we simply linearly extrapolate the absolute magnitude in each band as a function of log mass for masses below 0.09\,\msun, with no dependence on age or metallicity. This is not very accurate, but is preferable to zero flux. For systems with primary masses above 0.09\,\msun, the exact extrapolation has little influence on the inferred masses. \changed{It turns out that none of the primaries in our Orbital300 sample (defined below) has a median mass below this limit.}

Summing the fluxes for both components in a given band gives the flux we would expect to measure
\begin{alignat}{2}
\hat{f}_X \,&=\, \hat{f}_{1,X} + \hat{f}_{2,X} \ .
\label{eqn:app_flux_sum}
\end{alignat}
Adopting a Gaussian noise model for the \gaia\ flux measurements and assuming the forward model is true, the probability of measuring a flux \flux{X} from an unresolved binary system is described by a Gaussian with mean $\hat{f}_X$ and standard deviation $\Sigma_{f_X}$.
The latter is
\begin{alignat}{2}
\Sigma^2_{f_X} \,&=\, \sigma_{f_X}^2 + 0.848 s_X^2 (\hat{f}_{1,X}^{\,2} + \hat{f}_{2,X}^{\,2})
\label{eqn:likelihood_flux_variance}
\end{alignat}
which is the quadrature combination of the uncertainty in the measured flux $\sigma_{f_X}$, and the uncertainties in the forward model fit for each component $s_X$ (in magnitudes). The factor $(\ln(10)/2.5)^2 = 0.848$ comes from the flux--to--magnitude conversion.
We set $s_X$ equal to the median absolute deviation of the forward model fits quoted above.

\subsection{The posterior probability density function}\label{sec:posterior_pdf}

The posterior PDF is a product of the prior over the parameters, $P(\theta)$, and the four likelihood terms:
\begin{alignat}{2}
  P(\theta \given A, \parallax, \{f_X\}) 
  \,&=\, P(\theta) \, P(A \given \theta) \, \prod_X P(f_X \given \theta, \parallax)
  \label{eqn:posteriorPDF}
\end{alignat}
where $A = \sma{p}^3/T^2$ is the astrometric measurement and
$\theta = (\mass{1}, \mass{2}, \age, \met)$ is the four sampling parameters.
The last term on the right side is the product of the three Gaussian likelihoods in the \gaia\ fluxes described in section~\ref{sec:forward_model} ($X=G, B, R$). The middle term is the probability of obtaining the astrometric measurement given the parameters. When the model and measurements are perfect these quantities are linked by equation~\ref{eqn:kepler3_photocentre}.
But the right side of equation~\ref{eqn:kepler3_photocentre} is model-predicted (from the parameters $\theta$) whereas the left side is measured.
Making the convenient (but not entirely accurate) assumption that deviations between these are Gaussian, the likelihood term $P(A \given \theta)$ is then a Gaussian distribution with mean equal to the right side of equation~\ref{eqn:kepler3_photocentre} and standard deviation $\sigma_A$, where
\begin{alignat}{2}
  \left(\frac{\sigma_A}{A}\right)^2 \,&=\,
  \left(\frac{3\sigma_{\sma{p}}}{\sma{p}}\right)^2 +
  \left(\frac{2\sigma_T}{T}\right)^2
\label{eqn:likelihood_Afunc_variance}
\end{alignat}
is obtained by combining the uncertainties in the SMA and the period in quadrature, which is reasonable for small fractional uncertainties.
The posterior is therefore the product of four one-dimensional Gaussian likelihoods and the prior.
As mass and age must be positive, the sampling is done over the logarithm of the masses and age (metallicity is already a logarithmic expression of abundance).

We use a separable prior in all four parameters.
The mass priors are uniform in log mass between 0.001\,\msun\ and the maximum of the forward model grid (1.61\,\msun), which is minimally informative (\citealt{2017pbi..book.....B}; section 5.3.1), and zero outside.
We later impose absolute magnitude and colour cuts on the data to exclude stars likely to be more massive than this upper limit.
The age prior corresponds to a constant star formation rate from the maximum age
$\age=\age_{\rm max}$ in the grid
to the present date $\age=0$, so is $P(\log\age) \propto \age(\age_{\rm max} - \age)$, and zero outside this range.
The prior in metallicity increases linearly from the lowest metallicity in the grid $\met_{\rm min}$, so is
$P(\met) \propto \met - \met_{\rm min}$, and is zero outside the grid. This linear increase (triangle distribution) could reflect a steady enrichment over time. Neither this metallicity prior nor the age prior is particularly realistic, but they are arguably more realistic that uniform log priors. The \gaia\ fluxes are only weakly sensitive to age and metallicity, so a more complex and informative prior is not justified.
\changed{We later demonstrate that adopting instead the least informative priors possible (uniform log) hardly changes our results.}

\subsection{Sampling and summarizing the posterior}\label{sec:posterior_pdf_sampling}

We sample the posterior using a Markov Chain Monte Carlo (MCMC) method (emcee; \citealt{2013PASP..125..306F}) with 32 samplers for 1000 iterations.
The initial samples are drawn uniformly (in the sampling parameters) over the full range of the grid. For each of the 32 samples the higher mass is assigned to the primary and the lower mass to the secondary.
This is done just to provide a clear initial distinction between the primary and the secondary.
During the sampling there is no constraint that one mass be higher than the other, and samples are never swapped. This is essential in order to retain detailed balance in the MCMC and to allow the mass posteriors to overlap.

The last 500 iterations of the sampling, thinned by a factor of five, represent the four-dimensional posterior.
We summarize each marginal posterior using three quantiles:
the median, the 15.87 percentile, and the 84.12 percentile.
These are named -- for the primary mass, for example -- 
$\mass{1,{\rm med}}$, $\mass{1, {\rm lo}}$, and $\mass{1, {\rm up}}$ respectively.
The median is taken as the point estimate.
The difference between the median and the 15.87 percentile is the lower $1\sigma$-like uncertainty (`lower error bar'),
and
the difference between the 84.13 percentile and the median is the upper $1\sigma$-like uncertainty  (`upper error bar').
If the median mass of the secondary is higher than the median mass of the primary, they are relabelled at this point, although this is rarely necessary.

\changed{Our Monte Carlo chains are relatively short, although the number of chains (walkers) compensates for this to some degree. As we report below, longer chains do not change the inferred masses by much.}

\subsection{Adding spectroscopic (radial velocity) information}\label{sec:data_spectroscopic}

In spectroscopic binaries we observe variations in the radial velocity of the brighter component due to its orbit around the barycentre (type SB1).
These measurements permit the determination of two more Thiele--Innes parameters
(published in \gdr{3}; \citealt{2025A&A...693A.124G}).
\changed{Combining these with the astrometric data, we can determine
the spectroscopic SMA \sma{s} 
(\citealt{2023A&A...674A...9H}, appendix B).
This quantity is related to the physical SMA via equation~\ref{eqn:kepler3_photocentre} again, but now with $\flux1/\flux2 = \infty$ (and so $1/(1 + \flux1/\flux2)=0$), because in an SB1 we only observe the motion of the primary.
We therefore} define the `spectroscopic measurement' as the left side of the following equation
\begin{alignat}{2}
  \left(\frac{\sma{s}}{\au}\right)^3 \left(\frac{\period}{\yr}\right)^{-2} \,&=\,
   \frac{\mass2^3}{(\mass1 + \mass2)^2}
\label{eqn:spectroscopic_measurement}
\end{alignat}
with masses again in solar units.

\subsection{Comparison statistics}\label{sec:comparison_statistics}

To quantify the overall difference between two different mass estimates for a set of stars,
we use two metrics. The first is the median absolute residual (MAR)
\begin{equation}
  {\rm MAR} = {\rm median}(| y_i - x_i |)
   \label{eqn:MAR}
\end{equation}
\changed{where $y_i$ and $x_i$ are the two mass estimates for object $i$.}
This is a measure of the scatter between the estimates.
The bias between the estimates is the median residual
\begin{equation}
  {\rm bias} = {\rm median}(y_i - x_i) \ .
  \label{eqn:bias}
\end{equation}
Note that the MAR includes the impact of a bias, in the sense that if $x$ and $y$ differ just by a large constant factor, both the bias and MAR will be large.
\changed{When comparing our results to those by other authors, $y$ are our results and $x$ are the comparison results.}

\section{Results using astrometry and Gaia photometry}\label{sec:results_main}

\begin{figure}[t]
  \begin{center}
  \includegraphics[width=0.49\textwidth, angle=0]{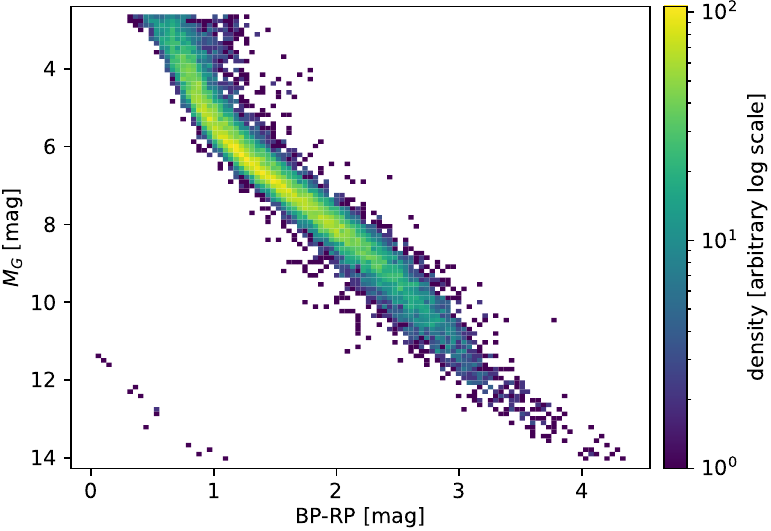}
  \caption{Colour--absolute magnitude diagram for the Orbital300 sample assuming zero extinction.
  \label{fig:input_data_CQD_nss_3_Orbital_3parsol_noNIR_20250916}
}
\end{center}
\end{figure}

We select all systems with {\tt nss\_solution\_type=`Orbital'} from the  {\tt nss\_two\_body\_orbit} table\footnote{\url{https://gea.esac.esa.int/archive/documentation/GDR3/Gaia_archive/chap_datamodel/sec_dm_non--single_stars_tables/ssec_dm_nss_two_body_orbit.html}} in \gdr{3} that lie within our \gaia\ photometry limits, have matches in the infrared 2MASS \citep{2006AJ....131.1163S} and AllWISE \citep{2010AJ....140.1868W,2011ApJ...731...53M} point source catalogues, and that lie within 300\,\pc\ of the Sun.
These \gaia\ photometry limits are the same as those we used to build the forward models in section~\ref{sec:forward_model}, with the absolute G-band magnitude computed neglecting extinction and with distance equal to inverse parallax.
We only use the infrared photometry later in section~\ref{sec:results_withNIR_noSpectro}.
This selection provides 20\,334 systems, which we refer to as the `Orbital300' sample.
Its CAMD is shown in figure~\ref{fig:input_data_CQD_nss_3_Orbital_3parsol_noNIR_20250916}.
The sample appears to consist almost entirely of main sequence stars (plus a few white dwarfs), although secondaries could be other stellar or substellar types.

\begin{figure}[t]
  \begin{center}
  \includegraphics[width=0.49\textwidth, angle=0]{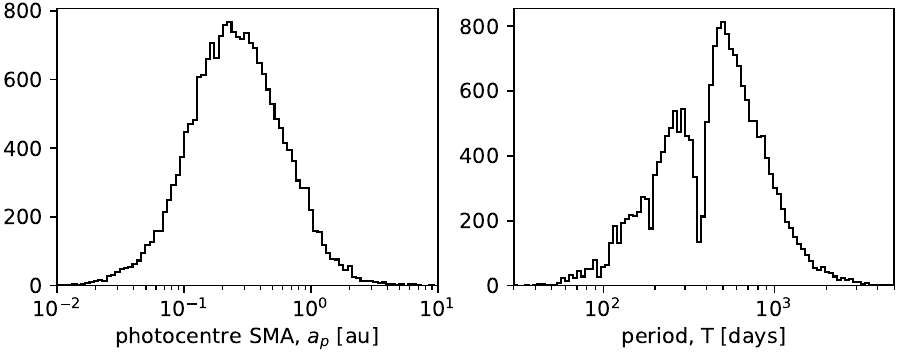}
  \includegraphics[width=0.49\textwidth, angle=0]{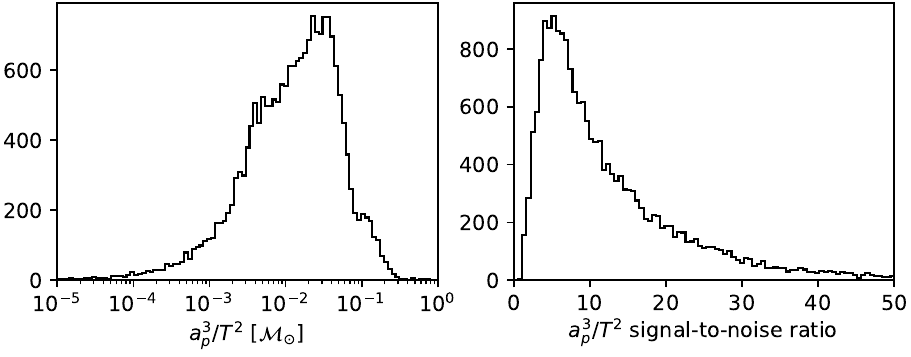}
  \includegraphics[width=0.49\textwidth, angle=0]{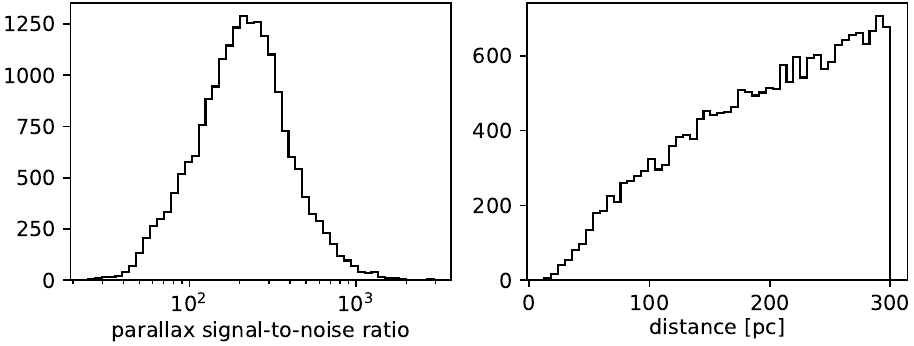}
  \caption{Distribution of the data for the Orbital300 sample.
    Top row: The photometric SMA $\sma{p}$ and period $T$.  Middle row: The astrometric measurement $\sma{p}^3/T^2$\changed{(with $\sma{p}$ in \au\ and $T$ in \yr, from the left side of equation~\ref{eqn:kepler3_photocentre})} (left) and its \snr\ (right).
  Bottom row: parallax \snr\ and distance (limited by construction to less than 300\,\pc).
  \label{fig:nss_3_Orbital_input_data}
}
\end{center}
\end{figure}

The distributions of the astrometric data are shown in figure~\ref{fig:nss_3_Orbital_input_data}.
The 1st and 99th percentiles of the period distribution are at 79 days and 2012 days (5.5 years) respectively.
Shorter periods are harder to detect, because they have smaller astrometric signal. Periods much longer than the data span of \gdr3, which is 1038 days (2.8 years), are similarly harder to detect.
The period distribution shows a sharp dip at around 365.25 days, which is the orbital period of \gaia\ around the Sun.
The \snr\ of the astrometric measurement $\sma{p}^3/T^2$ ranges from 1.8 and 55.8 (1st and 99th percentiles), with a median of just 9.4.
The median parallax \snr\ is 210; 99\% of sources have a parallax \snr\ above 49.

\begin{figure}[t]
\begin{center}
\includegraphics[width=0.49\textwidth, angle=0]{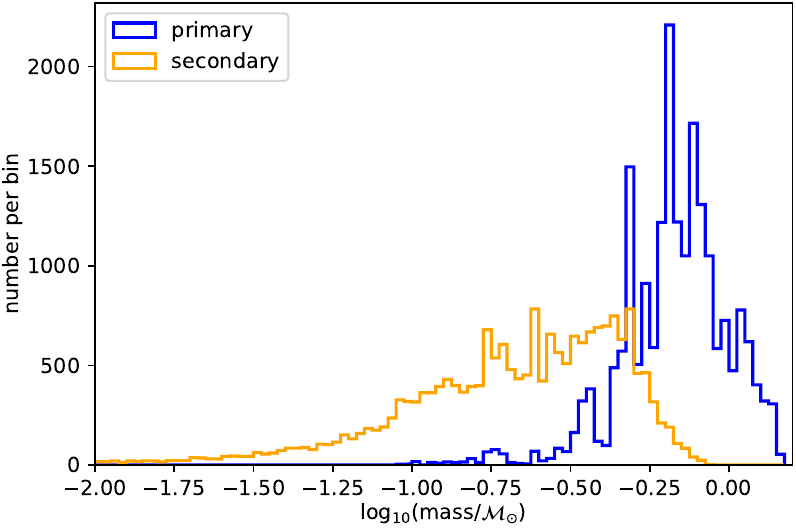}
\includegraphics[width=0.49\textwidth, angle=0]{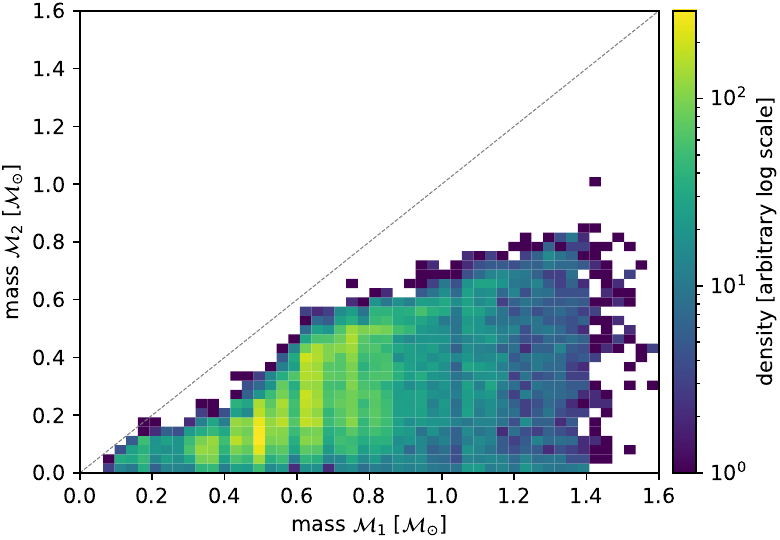}
\includegraphics[width=0.49\textwidth, angle=0]{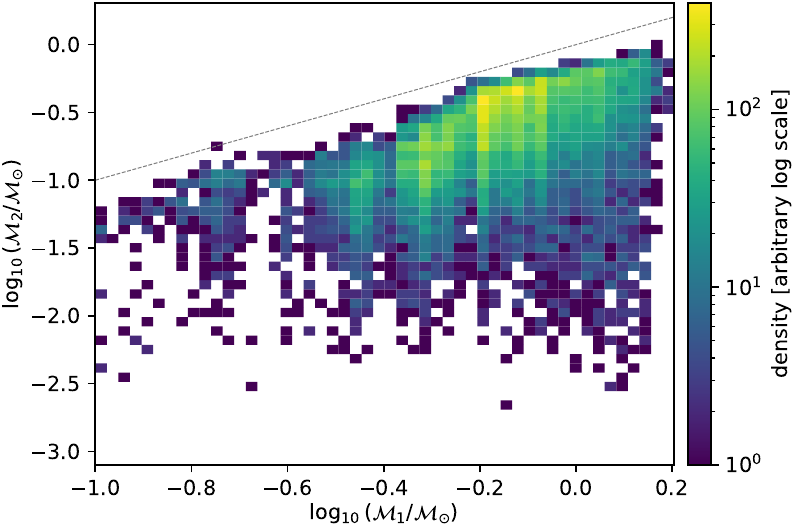}
\caption{Inferred masses (median of the posteriors) for the Orbital300 sample for the primary (\mass1, top) and secondary (\mass2, bottom).
\label{fig:nss_3_Orbital_mass_distribution}
}
\end{center}
\end{figure}

We run our parameter inference and summarize the posteriors as described in the previous section.
The distribution of the median of the inferred masses is shown in figure~\ref{fig:nss_3_Orbital_mass_distribution}.
The median of the mass distribution is 0.44\,\msun\ for the primaries and 0.07\,\msun\ for the secondaries.
There is a weak positive correlation between the primary and secondary masses.
\changed{The lowest median primary mass is 0.093\,\msun\ and the lowest median secondary mass is 0.0023\,\msun\ (the limit set by the prior is 0.001\,\msun\ in both cases). We discuss the lowest mass secondaries in section~\ref{sec:exoplanets}.}

\begin{figure}[t]
  \begin{center}
\includegraphics[width=0.49\textwidth, angle=0]{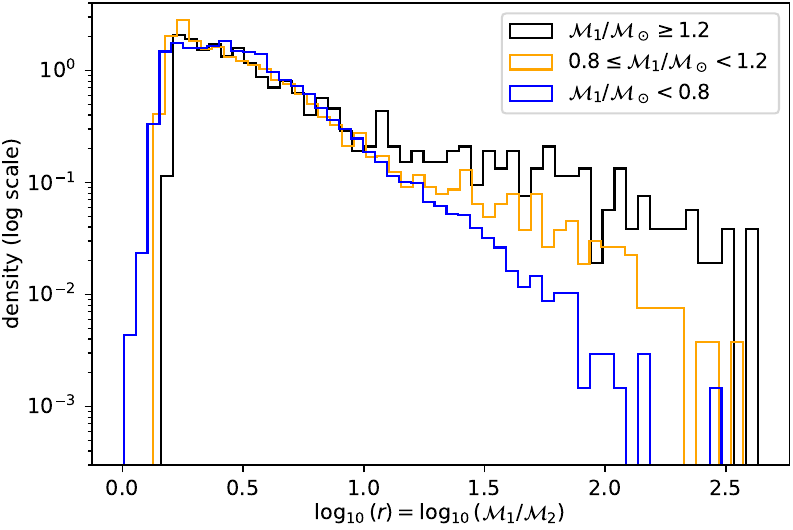}
\caption{Distribution of the mass ratio $r=\mass1/\mass2$ for the Orbital300 sample for three different primary mass ranges.
\label{fig:nss_3_Orbital_mass_ratio_distribution}
}
\end{center}
\end{figure}

Figure~\ref{fig:nss_3_Orbital_mass_ratio_distribution} shows the distribution of the mass ratio for three different primary mass ranges.
As we are particularly interested in the lowest mass companions, we use the parameter $r=\mass1/\mass2$ for the mass ratio, rather than its inverse. (For the log mass ratio this simply changes the sign.)
The peak of $\log_{10}r$ is at 0.2--0.4, corresponding to mass ratios of 1.6--2.5.
(Note that this is not the maximum of $r$, because a maximum is not invariant under a log transformation.)
Thus for lower mass primaries we do not see as many large mass ratios. This could be in part a consequence of the lower sensitivity to lower mass secondaries.

We refrain from fitting models to the mass ratio distribution, or from inferring population statistics, because the NSS sample suffers from various selection effects, as emphasised by \cite{2023A&A...674A...9H}.
It is notable, however, that for mass ratios below about 1.6, there is a steep drop in the number of binaries, meaning we find few near-equal mass binaries.
\gaia's high precision astrometry makes it suitable for identifying high mass ratio systems. Nonetheless, 
the larger the mass ratio, the smaller the astrometric measurement (equation~\ref{eqn:kepler3_photocentre}) in general, and so the more difficult it is to detect high mass ratio systems. This observational selection effect could explain some of the slope in figure~\ref{fig:nss_3_Orbital_mass_ratio_distribution}.

\begin{figure}[t]
  \begin{center}
\includegraphics[width=0.49\textwidth, angle=0]{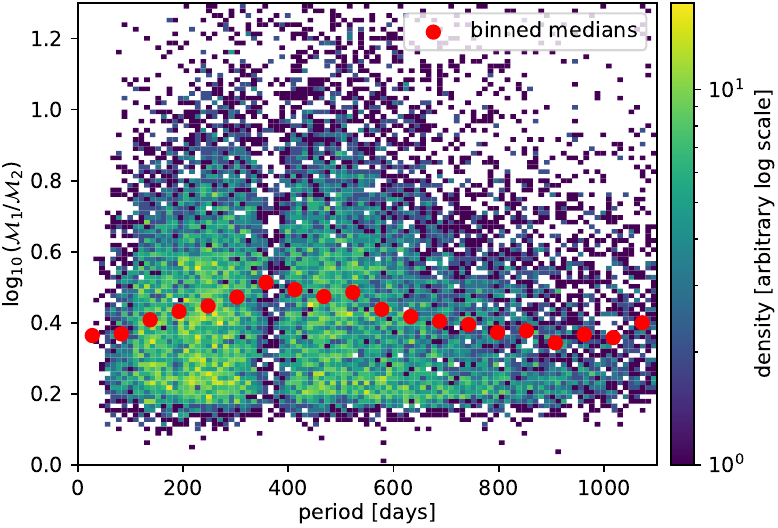}
\caption{Variation of the mass ratio $\mass1/\mass2$ with period for the Orbital300 sample. The red points are the median values of the mass ratio in equally-spaced period bins.
\label{fig:nss_3_Orbital_massratio_vs_period}
}
\end{center}
\end{figure}

The variation of the mass ratio with the orbital period is shown in figure~\ref{fig:nss_3_Orbital_massratio_vs_period}. We see a slight increase in the median mass ratio at intermediate periods.

\begin{figure}[t]
\begin{center}
  \includegraphics[width=0.49\textwidth, angle=0]{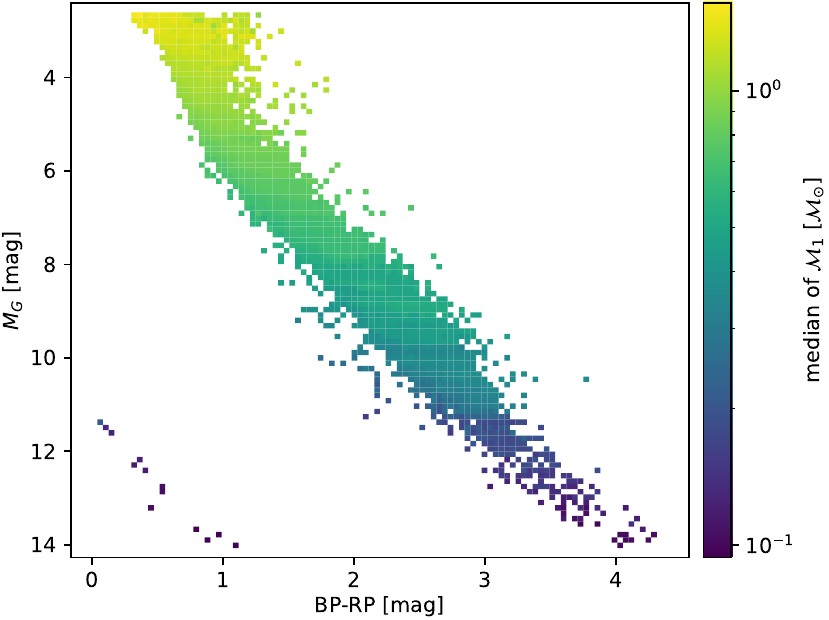}
  \includegraphics[width=0.49\textwidth, angle=0]{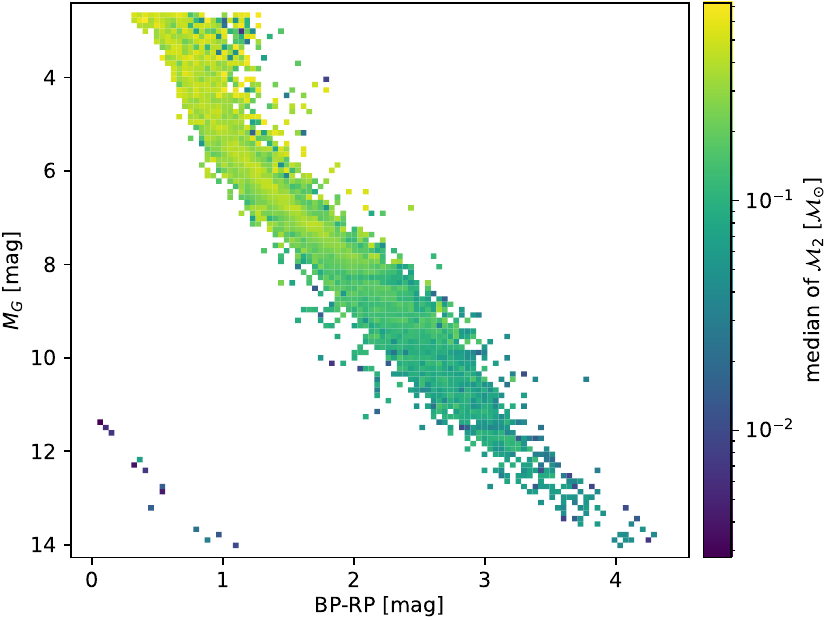}
\caption{System colour--absolute magnitude diagram for the Orbital300 sample colour-coded by the primary mass (top) and secondary mass (bottom). 
  \label{fig:nss_3_Orbital_CQD_mass_colourcoded}
  }
\end{center}
\end{figure}

The CAMD of the sample colour-coded by the median component masses is shown in figure~\ref{fig:nss_3_Orbital_CQD_mass_colourcoded}. The median over a set of stars is shown at each point. In both panels the magnitude and colour are those observed for the sytem. In the upper panel the primary mass varies fairly smoothly and continuously, with the mass decreasing for fainter and redder systems. This is what we expect when the primary dominates the system photometry. The colour-coding by the secondary mass in the lower panel shows a roughly similar trend.
But for systems brighter than $M_G \simeq 8.0$\,mag we see a locus of systems in the central part of the main sequence with higher average secondary masses, and thus lower mass ratios.

\begin{figure}[t]
  \begin{center}
\includegraphics[width=0.49\textwidth, angle=0]{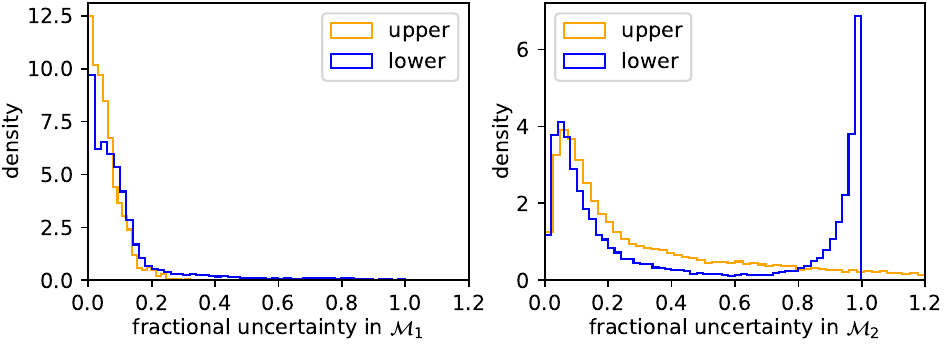}
\caption{Distribution of the inferred fractional $1\sigma$-like uncertainties in the masses for the Orbital300 sample for the primaries (left panel) and secondaries (right panel).
The lower fractional uncertainty cannot be larger than 1.0, whereas the upper one is unbounded.
There is a small number of secondaries with upper fractional uncertainties beyond the limit of the plot.
  \label{fig:nss_3_Orbital_mass_fracunc}
}
\end{center}
\end{figure}

So far we have only considered the point mass estimates, which are medians of the posterior distributions.
However, the posteriors can have significant widths, something we characterize
with the lower and upper $1\sigma$-like uncertainties (see section~\ref{sec:posterior_pdf_sampling}). Dividing each of these two widths by the median gives a lower and an upper fractional uncertainty respectively.
These are plotted for the primary and secondary mass estimates in figure~\ref{fig:nss_3_Orbital_mass_fracunc}.
Note that the lower fractional uncertainty cannot be larger than 1.0, whereas the upper one is unbounded.
The uncertainties for the primaries are relatively small,
with median lower and upper fractional uncertainties of 0.07 and 0.05 respectively.
90\% have lower and upper fractional uncertainties less than 0.23 and 0.12 respectively.
The fractional uncertainties for the secondaries, in contrast, are often much larger; only 50\% have lower and upper fractional uncertainties less than 0.25. A sizeable fraction have lower fractional uncertainties that are as nearly as large as 1.0. This means that the posterior permits an arbitrarily low mass (down to 0.001\,\msun, the lower limit of the inference). There is no obvious trend of the size of this fractional uncertainty with the size of the astrometric measurement.

\begin{figure*}[t]
  \begin{center}
\includegraphics[width=0.99\textwidth, angle=0]{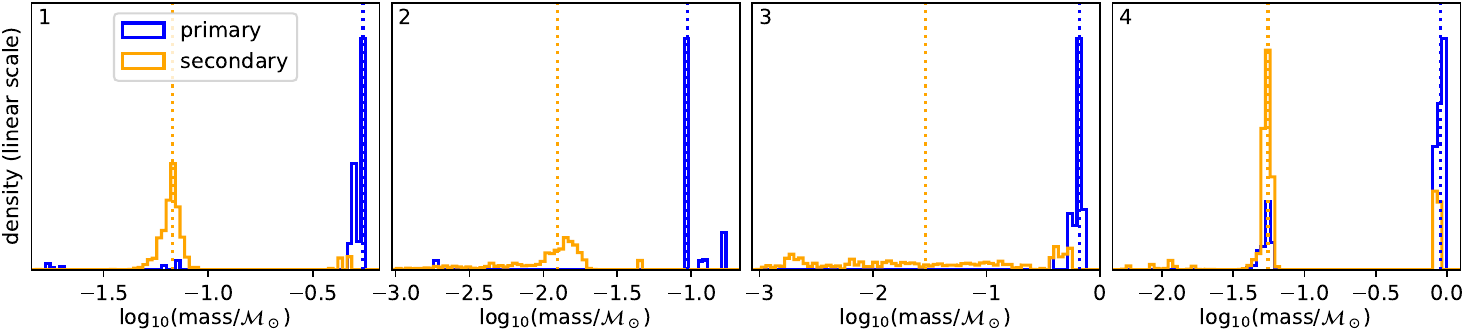}
\caption{Examples of the marginalized mass posteriors for the primary (blue) and secondary (orange) for \changed{four sources (one per panel).
Panel 1 is typical, panels 2 and 3 show examples of long tails to low masses for the secondaries, and panel 4 shows where some of the MCMC chains moved between primary and secondary during the sampling.}
The vertical dotted lines show the median masses.
  \label{fig:mass_posterior_examples}
}
\end{center}
\end{figure*}

Some example posteriors for the primaries and the secondaries are plotted as histograms in figure~\ref{fig:mass_posterior_examples}.
\changed{The first panel shows a typical case.}
It is quite common for a small fraction of the MCMC samples to be `disconnected' from the main set of samples when plotted as a histogram in this way. But the lower `peaks' are often either near to the main one or contain few samples.
Significant bimodality is very rare in the full sample. We do, however, see a few cases where there is a long low tail to low masses for the secondary, such as in panels 2 and 3 of figure~\ref{fig:mass_posterior_examples}. Panel 4 shows an example where some MCMC chains moved between primary and secondary during the sampling. However, the number of samples attributed to the other component is small, so the median is unaffected.

\changed{To test the impact of the prior on our results (section~\ref{sec:posterior_pdf}), we reran the inference using uniform log priors on age and metallicity, which are minimally informative but also less physical. Averaged over the sample, this increases the primary masses by 0.007\% and the secondary masses by 0.1\% (equation~\ref{eqn:bias}), which is negligible. There is a slightly larger scatter (equation~\ref{eqn:MAR}) of 1\% for primaries and 3\% for secondaries -- meaning a single source selected at random would differ by these amounts on average -- but these differences are much less than the widths of the posteriors, so are accounted for by the error bars.}

\changed{As MCMC is a stochastic process, the posteriors and their summaries vary if the MCMC is extended for more iterations or rerun with a different intialization. When we extend the MCMC to five times as many iterations (also for the burn-in), we find that the mass estimates change by only 1.5\% for the primaries and 3.9\% for the secondaries (scatter; equation~\ref{eqn:MAR}).
Rerunning with a different initialization causes less scatter.
Convergence noise therefore has only a small impact on our results.}

We provide our mass estimates as an electronic table with the entries show in Table~\ref{tab:mass_catalogue}.
The log flux ratio is the ratio computed from the forward model-predicted G-band fluxes (equation~\ref{eqn:forward_model})
at the median inferred values of the masses, age, and metallicity for each system.
This flux ratio is not equal to the value computed from equation~\ref{eqn:kepler3_photocentre} using the median masses and measured $\sma{p}$ and $T$, because the median stellar parameters are found by posterior sampling, and not solving that equation.

\begin{table*}
\begin{center}
\caption{Content of the catalogue of inferred masses for the Orbital300 sample.
\label{tab:mass_catalogue}
}
\small{
\begin{tabular}{rrrrrrrrrrrr}
  \hline
  \gaia\ DR3 {\tt source\_id} & $\mass{1,{\rm med}}$ & $\mass{1, {\rm lo}}$ & $\mass{1, {\rm up}}$ & $\mass{2, {\rm med}}$ & $\mass{2, {\rm lo}}$ & $\mass{2, {\rm up}}$ & $\sma{p}$ & $\sigma(\sma{p})$ & $T$ & $\sigma(T)$ & $\log_{\rm 10}(f_1/f_2)$ \\
    & [\msun] & [\msun] & [\msun] & [\msun] & [\msun] & [\msun] & [\au] & [\au] & [days] & [days] & \\
  \hline
1729398647131392 & 0.5001 & 0.4867 & 0.5493 & 0.1170 & 0.0035 & 0.1753 & 0.3118 & 0.0043 & 538.083 & 8.575 & 1.9576 \\
2488955023504768 & 0.4488 & 0.4248 & 0.4525 & 0.0775 & 0.0694 & 0.0878 & 0.2136 & 0.0154 & 441.589 & 4.821 & 2.1656 \\
3205080690546176 & 0.8473 & 0.7602 & 0.8642 & 0.0325 & 0.0035 & 0.5583 & 0.4405 & 0.0127 & 879.223 & 64.436 & 4.9103 \\
3679313799492864 & 0.3648 & 0.3588 & 0.3882 & 0.1232 & 0.0811 & 0.1259 & 0.0897 & 0.0035 & 1074.243 & 56.782 & 1.1449 \\
4016657710743168 & 1.4086 & 1.1721 & 1.5381 & 0.4151 & 0.0132 & 0.4474 & 0.4196 & 0.0192 & 822.289 & 26.599 & 2.7579 \\
5168082607675520 & 0.6835 & 0.6479 & 0.7370 & 0.3085 & 0.2906 & 0.4449 & 0.1792 & 0.0051 & 711.032 & 4.757 & 1.5329 \\
7326187710462720 & 1.0944 & 0.8926 & 1.1955 & 0.2941 & 0.0141 & 0.3830 & 0.1891 & 0.0035 & 300.369 & 1.320 & 2.5974 \\
7623811764175616 & 0.7208 & 0.7176 & 0.7258 & 0.1176 & 0.1135 & 0.1241 & 0.1723 & 0.0180 & 241.112 & 1.492 & 2.7660 \\
8499499760858368 & 0.8426 & 0.7643 & 0.8837 & 0.4637 & 0.2747 & 0.5320 & 0.2381 & 0.0094 & 448.716 & 0.957 & 1.4428 \\
8628662312765440 & 0.6880 & 0.6622 & 0.7278 & 0.1152 & 0.0038 & 0.4162 & 0.1884 & 0.0067 & 1243.326 & 33.863 & 2.7783 \\
  \hline
\end{tabular}
}
\tablefoot{
$\mass{1, {\rm med}}$ is the median mass of the primary, $\mass{1, {\rm lo}}$ and $\mass{1, {\rm up}}$ are the 16th and 84th percentiles of the mass posterior, representing the lower and upper uncertainties respectively; and likewise for the secondary, \mass2.
The table also includes the photometric SMA $\sma{p}$, the period $T$, and their $1\sigma$ uncertainties. The final column is log base 10 of the G-band flux ratio at the median parameters.
The electronic version of this table lists all 20\,334 systems sorted by increasing {\tt source\_id}, with fields reported to full precision (no rounding). There are no missing data.
 }
\end{center}
\end{table*}

There is an obvious set of 14 white dwarfs (primaries) in the bottom left corner of the CAMD of figure~\ref{fig:input_data_CQD_nss_3_Orbital_3parsol_noNIR_20250916}. Our photometric forward models are not appropriate for these stars, so we do not expect to infer appropriate masses. Indeed, we get median primary and secondary masses of around 0.1\,\msun\ and 0.008\,\msun\ respectively for these systems. We have left them in the published catalogue because they were not excluded from our analysis (they have negligible impact on the later comparison statistics).

\subsection{Relevance of flux ratio on the mass estimates}

\begin{figure}[t]
\begin{center}
\includegraphics[width=0.49\textwidth, angle=0]{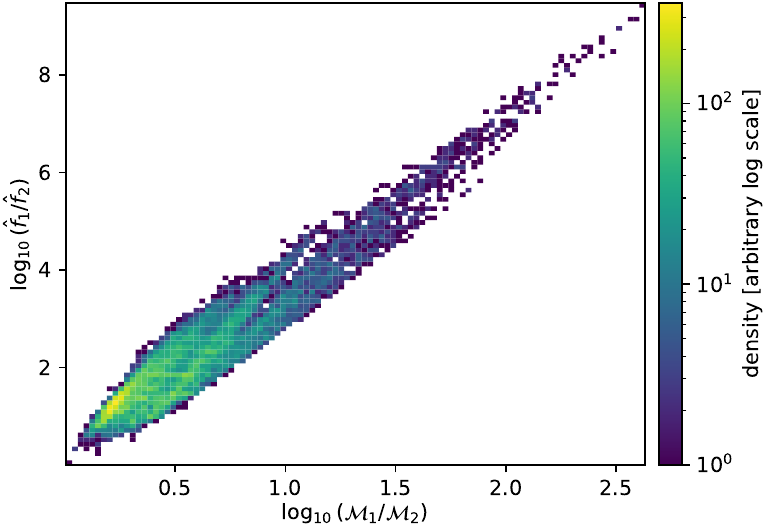}
\caption{Predicted G-band flux ratio vs inferred mass ratio for the Orbital300 sample.
  \label{fig:nss_3_Orbital_fluxratio_massratio}
  }
\end{center}
\end{figure}

Our forward model (equation~\ref{eqn:forward_model}) predicts the G-band flux for each component of the unresolved binary from the mass, age, and metallicity.
From this we compute the flux ratio at the median inferred values of the parameters.
Figure~\ref{fig:nss_3_Orbital_fluxratio_massratio} shows how this flux ratio varies with the inferred mass ratio for the Orbital300 sample. This is broadly what we expect for stellar binaries -- the flux ratio increasing with mass ratio, plus additional variance due to metallicity and age variations -- and is a consequence of our forward model. 

\begin{figure}[t]
\begin{center}
\includegraphics[width=0.49\textwidth, angle=0]{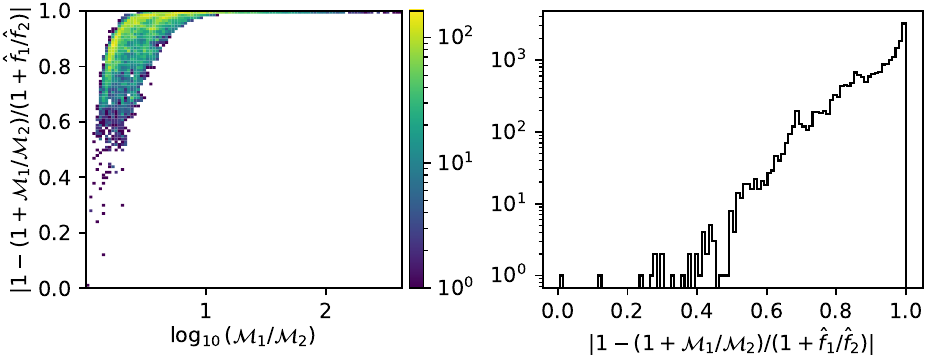}
\caption{Flux relevance ratio (equation~\ref{eqn:kepler3_photocentre}) plotted against the inferred mass ratio (left) for the Orbital300 sample shown on a logarithmic density scale. The right panel shows the distribution of the flux relevance ratio.
  \label{fig:nss_3_Orbital_frr}
  }
\end{center}
\end{figure}

The impact of the flux from the secondary on the determination of the masses via the astrometry is determined by the size of the cubed quantity on the right side of equation~\ref{eqn:kepler3_photocentre}, or more conveniently by 
\begin{alignat}{2}
 \frr \,&=\,
  \left| 1 - \frac{1 + \mass1/\mass2}{1 + \hat{f}_1/\hat{f}_2} \right| \ .
\label{eqn:frr}
\end{alignat}
which we call the `flux relevance ratio'. When the secondary contributes negligible flux, such that the flux ratio $\hat{f}_1/\hat{f}_2$ is very large, then \frr\ is close to one. When, instead, this flux ratio is closer to unity for a given mass ratio, then \frr\ is less than one, and the smaller \frr, the more relevant the flux is from the secondary in the interpretation of the astrometric measurement.
The distribution of \frr\ is shown in figure~\ref{fig:nss_3_Orbital_frr}. The majority of systems have \frr\ near to unity, but there is tail of systems with \frr\ as low as 0.4, and a handful at even lower values. The figure also shows how \frr\ varies with the primary mass. Unsurprisingly, it is larger for lower primary masses, because this is where the effect of the secondary on both the astrometry and photometry is more noticeable.

We test the importance of accounting for the flux of the secondary more directly by repeating the inference with this flux fixed to zero. In this case $f_2=0$ both in equation~\ref{eqn:kepler3_photocentre} and when computing the photometric likelihoods. We find that the estimated masses of the secondaries are then 9\% lower on average (median), and 10\% of the sources have masses that are 40\% or more lower.
This confirms that we need to accommodate for the unknown flux contribution of the secondary when trying to estimate its mass.

\subsection{Correction for interstellar extinction}\label{results:extinction_correction}

So far we have ignored interstellar extinction. Such extinction will dim and redden the observed photometry of some stars to some degree. In reality these stars are therefore brighter and bluer, and so more massive than what we infer.
Individual extinction estimates for some of our target stars are available in \gdr3\ from the GSP-Phot algorithm \cite{2023A&A...674A..27A,2023A&A...674A..28F}. This algorithm estimated the extinction parameter $A_0$ self-consistently together with the intrinsic stellar parameters via stellar models, and from this and the photometric passbands computed the extinction in each of the G, BP, and RP bands.
Two-thirds of our Orbital300 sample -- 14\,173 sources -- have extinction estimates, of which 13\,710 have extinction-corrected magnitudes that lie within the bounds of our forward model.
Their median G-band extinction $A_G$ is 0.21\,mag (10th and 90th percentiles are 0.004 and 0.56\,mag).
The GSP-Phot authors note that their extinction estimates may not be very accurate; in particular some may be overestimated.
However, these extinction values seem plausible for sources over the whole sky within 300\,\pc, and so we can use them at least to assess the impact of neglected extinction.

We use the GSP-Phot extinctions to adjust the forward-model predicted fluxes in equation~\ref{eqn:apparent_flux_from_absmag} for all three bands during the inference procedure. We then compare the inferred masses (posterior median) source-by-source with the extinction-free masses inferred earlier. When including the extinction, the primary masses are estimated to be 2.2\% higher on average (median), and the secondary masses 0.4\% higher. The $\mass1/\mass2$ ratio is 1.6\% higher. Hence there is very little systematic increase in the mass estimates when including plausible extinctions.
The scatter between the estimates is somewhat larger than these biases,
indicating that some individual stars' mass estimates vary by more, although some of this is also MCMC convergence noise.
Overall we can say that for stars within 300\,\pc, neglecting extinction has very little impact on the mass estimates with our method.

\section{Comparison to other estimates}\label{sec:comparison}

As part of the \gdr3\ validation,
\cite{2023A&A...674A..34G} examined various combinations of \gaia\ NSS data to estimate the mass of the primary and/or secondary. Which data are used to estimate which masses is indicated by the flag {\tt combination\_method} in their table.\footnote{\url{https://gea.esac.esa.int/archive/documentation/GDR3/Gaia_archive/chap_datamodel/sec_dm_performance_verification/ssec_dm_binary_masses.html}}
In several cases they had to adopt a value for the mass of the primary in order to estimate the mass of the secondary. Often this primary mass was estimated using \gaia\ photometry and isochrones. This is similar to what we are doing in the present study, except that we estimate the masses of both components simultaneously using the astrometry and photometry.
Here we refer to the mass estimates from \cite{2023A&A...674A..34G} as the `comparison set'. Some of their \mass1\ masses are lower than their \mass2\ masses,
which we then swap to be consistent with our results
(our primary always has the higher median mass).
For the Orbital300 sample analysed in the previous section,
the comparison set provides estimates of the primary mass in 19\,603 systems but of the secondary mass in only 1574 systems, the latter being a subset of the former. (They have more cases where the secondary has a lower and/or upper mass limit, but we don't use those here.)

\begin{figure}[t]
\begin{center}
\includegraphics[width=0.49\textwidth, angle=0]{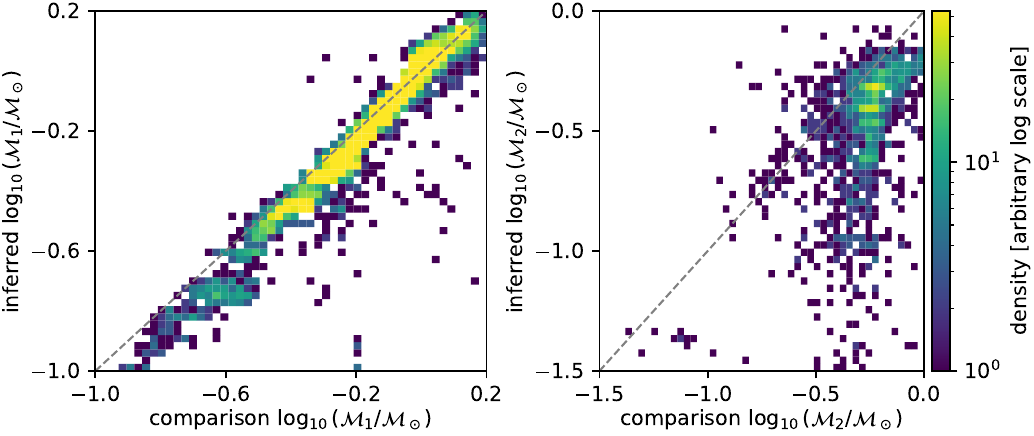}
\caption{Comparison of our mass estimates on the Orbital300 sample using astrometry and photometry
with those from \cite{2023A&A...674A..34G} who adopt a mass for the primary then infer the mass of the secondary from a variety of data (in many cases there is no secondary mass estimate).
The two panels show the primaries (left, 19\,603 stars) and secondaries (right, 1574 stars).
A few sources lie outside the range of the plots.
\label{fig:masses_comparison_2D_histogram_nss_3_Orbital}
}
\end{center}

\end{figure}

Figure~\ref{fig:masses_comparison_2D_histogram_nss_3_Orbital} compares the two mass estimates for the primaries and secondaries. The agreement for the primaries is reasonable. In contrast, we see some significant differences for many of the secondaries, with our estimates extending to lower masses than the comparison ones. Recall that our estimates are the medians of a posterior which for the secondaries can have a long tail to low masses, as shown earlier in figure~\ref{fig:nss_3_Orbital_mass_fracunc}.

\begin{figure}[t]
\begin{center}
\includegraphics[width=0.49\textwidth, angle=0]{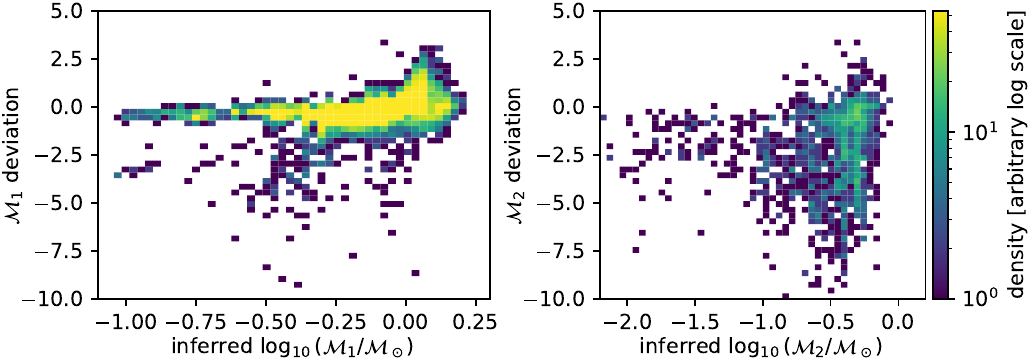}
\caption{Deviations, as defined by equation~\ref{eqn:deviation}, for the results shown in Figure~\ref{fig:masses_comparison_2D_histogram_nss_3_Orbital}.
\label{fig:masses_comparison_deviation_nss_3_Orbital}
}
\end{center}
\end{figure}

\changed{Figure~\ref{fig:nss_3_Orbital_mass_fracunc} compares the median mass estimates, yet the uncertainties (posterior widths) in both estimates are non-negligible. To accommodate the uncertainties in a comparison, we define the `deviation' as the difference between the mass estimates divided by their combined error bar, using the lower error bar (subscript `lo') of one estimate and the upper error bar (subscript `up') of the other, according to which estimate is higher (to get the `neighbouring' error bar):
\begin{alignat}{2}
  \textrm{deviation}\,=\,
  \begin{cases}
    \frac{\displaystyle y - x}{\displaystyle \sqrt{\sigma_{y,lo}^2 + \sigma_{x,up}^2}} & \:{\rm if}~~ \displaystyle y - x \geq 0 \\ \\
    \frac{\displaystyle y - x}{\displaystyle \sqrt{\sigma_{y,up}^2 + \sigma_{x,lo}^2}} & \:{\rm otherwise} \ .
  \end{cases}
  \label{eqn:deviation}
\end{alignat}
These deviations -- with $y$ as our estimates and $x$ as the comparisons -- are plotted against our estimates in Figure~\ref{fig:masses_comparison_deviation_nss_3_Orbital}. While the deviations for the primary masses are mostly just `one or two sigma', in some cases the deviations for the secondary masses are considerably larger, showing that some differences are statistically significant.}

\changed{We quantify the difference in the mass estimates using the statistics in section~\ref{sec:comparison_statistics}.}
For the primaries, the bias and scatter (MAR) in $\log_{10}\mass1$ are $-0.024$\,dex and $0.028$\,dex respectively. Differences in log base 10 quantities correspond to ratios 2.3 times larger, so we can also interpret these values as a bias of $-5.5$\% and a MAR of $6.4$\% in \mass1.
In absolute terms, the bias is $-0.037$\,\msun\ and the MAR is $0.044$\,\msun.
The comparison set does include a few very high mass estimates (up to 123\,\msun), but as our statistics are based on the median, these have little impact.

For the secondaries, the bias and MAR in $\log_{10}\mass2$ are $-0.16$\,dex ($-38$\%) and $0.17$\,dex (39\%) respectively.
In this case the MAR is dominated by the bias: the MAR around the bias is $0.11$\,dex.
This systematic difference is unlikely to be caused by us neglecting extinction
because that should equally effect the primary mass estimates.
In absolute terms, the bias and MAR are $-0.18$\,\msun\ and $0.18$\,\msun\ respectively, although these values make little sense given that \mass2\ spans two orders of magnitude (see figure~\ref{fig:nss_3_Orbital_mass_distribution}).

Among the primary masses compared, 96\% have the flag {\tt combination\_method}
in the comparison table equal to {\tt Orbital+M1}.
This means that \mass1\ has been estimated from Gaia photometry.
Among the secondary masses, 53\% have {\tt combination\_method} set to {\tt Orbital+M1}, 
which means that \mass2\ has been estimated from the photometrically-inferred \mass1\ together with the astrometric measurement (the astrometric orbit).
The bias and MAR of our $\log_{10}\mass2$ inferences compared to this subset are higher than for the complete set, at $-0.22$\,dex and $0.22$\,dex respectively.
The complementary subset -- the other 47\% -- contains mostly solutions with {\tt combination\_method} equal to
{\tt Orbital+SB1+M1}.
The bias and MAR of our $\log_{10}\mass2$ inferences for these are lower, at $-0.10$\,dex and $0.13$\,dex respectively.
The comparison mass estimates in this complementary subset have used the spectroscopic orbit in addition to the astrometric orbit (and photometry), so they should be more accurate than the other comparison mass estimates.
Our inferences, in contrast, have just used the photometry and astrometry in both subsets.

\begin{figure}[t]
\begin{center}
\includegraphics[width=0.49\textwidth, angle=0]{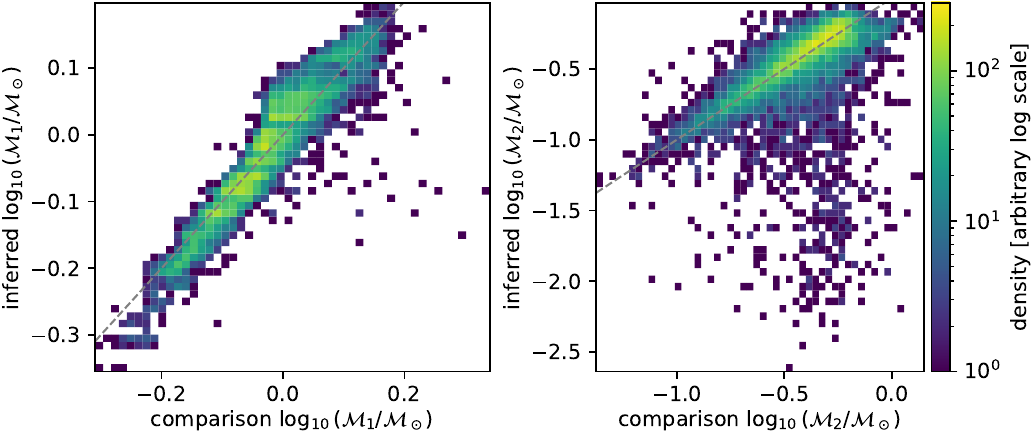}
\caption{Comparison of our mass estimates on the AstroSpectro300 sample when not using the spectroscopic measurement with mass estimates 
  from \cite{2023A&A...674A..34G} who adopt a primary mass and then used the astrometry and the spectroscopy to estimate the secondary mass. The two panels show the primaries (left) and secondaries (right), both with 9674 stars.  The horizontal and vertical axis do not span a common range in each panel.
\label{fig:masses_comparison_2D_histogram_nss_4_AstroSpectroSB1+M1_noNIR_noSpectro}
}
\end{center}
\end{figure}

\begin{figure}[t]
\begin{center}
\includegraphics[width=0.49\textwidth, angle=0]{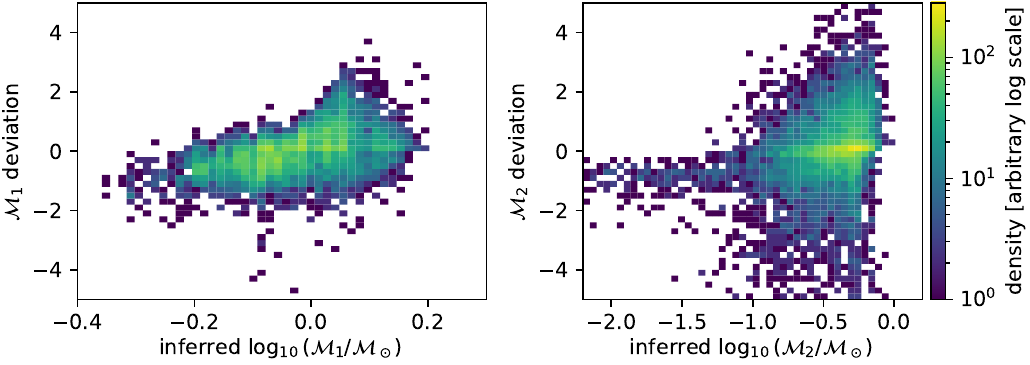}
\caption{Deviations, as defined in equation~\ref{eqn:deviation}, for the results shown in Figure~\ref{fig:masses_comparison_2D_histogram_nss_4_AstroSpectroSB1+M1_noNIR_noSpectro}.
\label{fig:masses_comparison_deviation_nss_4_AstroSpectroSB1+M1_noNIR_noSpectro}
}
\end{center}
\end{figure}

To investigate this further, we repeat the comparison on a different sample, namely those that have {\tt combination\_method} set to {\tt AstroSpectroSB1+M1} and that lie within 300\,\pc, a set of 9674 systems we call the AstroSpectro300 sample.
(This sample is defined in section~\ref{sec:results_noNIR_withSpectro} below.)
These are all single-lined spectroscopic binaries, and so \cite{2023A&A...674A..34G} used the combined spectroscopic and astrometric orbit along with an adopted value of \mass1\ to estimate \mass2. We infer the masses with our method using just the astrometry and photometry (but not the spectroscopy) and compare results.
The bias and MAR for $\log_{10}\mass1$ are $-0.001$\,dex and $0.019$ respectively, and for  $\log_{10}\mass2$ are $+0.007$\,dex and $0.062$\,dex respectively.
This is a much better agreement than we had above, as can be see in figure~\ref{fig:masses_comparison_2D_histogram_nss_4_AstroSpectroSB1+M1_noNIR_noSpectro}, \changed{which compares the masses, and
figure~\ref{fig:masses_comparison_deviation_nss_4_AstroSpectroSB1+M1_noNIR_noSpectro}, which shows the deviations}.
Part of the reason for the better agreement may be that the AstroSpectro300 sample is brighter -- compare figures~\ref{fig:input_data_CQD_nss_3_Orbital_3parsol_noNIR_20250916} and~\ref{fig:input_data_CQD_nss_4_AstroSpectroSB1+M1_withSpectro_noNIR_20250915} -- implying better quality photometry and astrometry than the Orbital300 sample.
Indeed, with the Orbital300 sample, there is a trend towards increasing discrepancy in the secondary masses toward fainter magnitudes. But there is no such trend with the \snr\ of the astrometric measurement.
In any case, the improved agreement with the AstroSpectro300 sample suggests that our method can get reasonable mass results for both components even without using the spectroscopic data.

\cite{2024A&A...682A..12P} also estimated the masses of astrometric binaries from \gdr3\ data, taking
advantage of the results from the MSC algorithm in \gdr3. MSC attempted to estimate the stellar parameters of both components of an unresolved binary assuming the target was a binary with a flux ratio less than five,
using the composite BP/RP spectrum (\citealt{2023A&A...674A..28F}). 
The validation undertaken by the DPAC showed that these stellar parameters are not very reliable. however, and should be used with caution (\citealt{2023A&A...674A..28F,2023A&A...674A..32B}).
Adopting a unique mass--luminosity relation,
\cite{2024A&A...682A..12P} derive the component masses from the astrometric measurement
and the system's G-band photometry -- corrected for extinction using MSC's estimate -- 
by adjusting the G-band magnitude difference between the components in discrete steps and selecting the one with the best agreement with the data.

A total of 10\,321 of their sources overlap with our Orbital300 sample, although 21\%
of their primary masses line in the range 10--90\,\msun, which seems an implausibly large fraction. Excluding these, we nonetheless find a good agreement between our primary mass estimates, with a bias of $+0.006$\,dex (1.4\%) and a MAR of $0.018$\,dex (4.2\%) in $\log_{10}\mass1$. The secondary masses show very poor agreement, however. Specifically, our secondary masses are invariably lower. Most of their systems have mass ratios ($\mass1/\mass2$) close to 1.0; 90\% have mass ratios below 1.5. Their method is simpler than ours, but also uses less photometric data and appears not to allow for the broad intrinsic scatter in the stellar mass--luminosity relation (e.g.\ figure~\ref{fig:forward_model_MG_vs_mass_met_colourcoded}; see the discussion in section~\ref{sec:breaking_degeneracy}).

\section{Planetary-mass companions}\label{sec:exoplanets}

The Orbital300 sample contains
\changed{155 systems in which the secondary has a median mass below 0.012\,\msun\ (about 13\,\mjup).
Many of these have posteriors extending to much higher masses, however, so they should not be taken as good planetary-mass candidates.
We instead identify the 22 systems where $\mass{2, {\rm up}}<0.012$\,\msun, i.e.\
the 84th percentile of their mass posterior (`median plus one sigma') is below 0.012\,\msun.}
Five of these systems lie on the white dwarf sequence, so their primary and secondary masses are unreliable. This leaves 17 objects that we refer to as exoplanet candidates. These are listed in Table~\ref{tab:exoplanet_candidates}.

\begin{table*}
\begin{center}
 \caption{All 17 exoplanet companion candidates, sorted by increasing $\mass{2, {\rm med}}$.
\label{tab:exoplanet_candidates}
}
\small{ 
\begin{tabular}{rrrrrrrrrrrr}
  \hline
  \gaia\ DR3 {\tt source\_id} & $\mass{1,{\rm med}}$ & $\mass{1, {\rm lo}}$ & $\mass{1, {\rm up}}$ & $\mass{2, {\rm med}}$ & $\mass{2, {\rm lo}}$ & $\mass{2, {\rm up}}$ & $\sma{p}$ & $\sigma(\sma{p})$ & $T$ & $\sigma(T)$ & $\log_{\rm 10}(f_1/f_2)$ \\
    & [\msun] & [\msun] & [\msun] & [\msun] & [\msun] & [\msun] & [\au] & [\au] & [days] & [days] & \\
  \hline
1768911697089736320 & 1.1110 & 1.0148 & 1.1776 & 0.0041 & 0.0017 & 0.0086 & 0.1415 & 0.0113 & 656.102 & 15.433 & 8.6627 \\
1462767459023424512 & 0.3362 & 0.2904 & 0.3438 & 0.0053 & 0.0021 & 0.0071 & 0.0235 & 0.0018 & 781.082 & 33.096 & 5.5126 \\
430892357759527424 & 0.1279 & 0.1239 & 0.1489 & 0.0055 & 0.0049 & 0.0062 & 0.1814 & 0.0156 & 495.280 & 8.581 & 4.7508 \\
6781298098147816192 & 0.3070 & 0.2388 & 0.3173 & 0.0058 & 0.0049 & 0.0070 & 0.1096 & 0.0031 & 602.376 & 11.731 & 5.2933 \\
4702845638429469056 & 0.3682 & 0.3408 & 0.4063 & 0.0058 & 0.0052 & 0.0065 & 0.0782 & 0.0030 & 596.613 & 12.481 & 5.8038 \\
4188996885011268608 & 0.1904 & 0.1850 & 0.1967 & 0.0065 & 0.0053 & 0.0072 & 0.0106 & 0.0008 & 406.363 & 3.501 & 4.6143 \\
4159075462792179456 & 0.3336 & 0.2892 & 0.3412 & 0.0069 & 0.0042 & 0.0083 & 0.3052 & 0.0063 & 823.125 & 97.043 & 5.0892 \\
4842246017566495232 & 0.3463 & 0.2882 & 0.3491 & 0.0072 & 0.0058 & 0.0079 & 0.0499 & 0.0012 & 465.430 & 6.461 & 5.2302 \\
6694115931396057728 & 0.2760 & 0.2466 & 0.2898 & 0.0077 & 0.0069 & 0.0083 & 0.0418 & 0.0004 & 459.346 & 2.421 & 5.0974 \\
557717892980808960 & 0.5347 & 0.5288 & 0.5384 & 0.0077 & 0.0069 & 0.0084 & 0.1982 & 0.0204 & 522.575 & 8.134 & 5.6757 \\
5796338299045711232 & 0.1290 & 0.1264 & 0.1505 & 0.0090 & 0.0084 & 0.0098 & 0.0359 & 0.0007 & 488.053 & 3.944 & 3.9641 \\
2845310284780420864 & 0.3509 & 0.3293 & 0.3567 & 0.0092 & 0.0085 & 0.0102 & 0.0269 & 0.0007 & 417.497 & 6.567 & 4.7233 \\
2277249663873880576 & 0.4257 & 0.4243 & 0.4675 & 0.0094 & 0.0086 & 0.0105 & 0.1205 & 0.0023 & 599.359 & 11.777 & 5.2529 \\
198464052134353536 & 0.1524 & 0.1521 & 0.2046 & 0.0094 & 0.0083 & 0.0112 & 0.0925 & 0.0007 & 301.518 & 2.699 & 4.0084 \\
5486916932205092352 & 0.3167 & 0.2424 & 0.3181 & 0.0101 & 0.0090 & 0.0106 & 0.0181 & 0.0006 & 253.479 & 0.926 & 4.5034 \\
2271703211828512896 & 0.1181 & 0.1172 & 0.1643 & 0.0101 & 0.0028 & 0.0119 & 0.0558 & 0.0010 & 622.841 & 10.586 & 3.6717 \\
5036787935627755520 & 0.1871 & 0.1828 & 0.1925 & 0.0104 & 0.0094 & 0.0113 & 0.0527 & 0.0014 & 850.942 & 79.281 & 3.9374 \\
  \hline
\end{tabular}
}
\tablefoot{
 These are the systems in the Orbital300 sample that have $\mass{2, {\rm up}}<0.012$\,\msun\ and are not on the white dwarf sequence of the colour--absolute magnitude diagram. Columns as in table~\ref{tab:mass_catalogue}.
}
\end{center}
\end{table*}

The \gaia\ Consortium has published a list of 72 `astrometric objects of interest' (ASOIs) that may be binaries with an exoplanet secondary.\footnote{\url{https://www.cosmos.esa.int/web/gaia/exoplanets}. The table list 73 ASOIs as of 2026-01-12, but one is a duplicate.}
Of these, 52 are in the Orbital300 sample.
Of those, 10 are in our exoplanet candidate list.

\cite{2025AJ....169..107S} followed up 28 of the ASOIs using radial velocities, of which five are in our exoplanet candidate list.
Of these, four are designated as SB2 by \cite{2025AJ....169..107S}. We would not expect to see the lines of an planetary-mass secondary in the spectrum, so this suggests these four objects are not planets (false positives in our list). A fifth object is designated `Single', meaning only one set of lines (the primary) was seen. The remaining five ASOIs are designated `Unknown', indicating there has been no known follow-up.

In total, \cite{2025AJ....169..107S} find 23 SBs. Of these, 22 are in our Orbital300 sample. As just noted, four are in our exoplanet candidate list. But 18 are not. Thus while our method may not be very accurate at estimating the masses of exoplanet companions -- and it is not designed to do so -- it shows some promise in identifying stellar or brown dwarf companions.
It can therefore help to clean exoplanet target lists for follow-up.

The above caveats notwithstanding, our mass estimates for one new \gaia\ exoplanet and one new \gaia\ brown dwarf identified by \cite{2025AJ....169..107S}  agree well with their estimates which used high-precision (10--15\,\mpers) multi-epoch radial velocities and multiband photometry in addition to the \gaia\ astrometry.
For Gaia-4b (GDR3 1457486023639239296) they report a mass of $11.8\pm0.7$\,\mjup, which compares to our estimate of $11.8^{+2.8}_{-8.2}$\,\mjup.
For Gaia-5b (GDR3 2074815898041643520) they report a mass of $20.9\pm0.7$\,\mjup, which compares to our estimate of $21.4^{+1.3}_{-1.9}$\,\mjup.
Neither of these is in our exoplanet candidate list because $\mass{2, {\rm up}}>0.12$\,\msun\ in both cases.
The other three \gaia\ objects with exoplanet designations, Gaia-1b, \mbox{-2b}, and \mbox{-3b} \citep{2022A&A...663A.101P,2023A&A...677L..15S} are not in our Orbital300 sample.

\section{Inclusion of infrared photometry}\label{sec:results_withNIR_noSpectro}

As we are using photometry to help estimate the masses of the two components, it is reasonable to expect that more photometric bands could improve the mass estimates, either in accuracy or in precision.
In this section we add the J and K$_s$ bands from 2MASS and the W1 and W2 bands from AllWISE and include these as four extra terms in the photometric likelihood (equation~\ref{eqn:posteriorPDF}). The astrometric likelihood is unchanged as we still use the \gaia\ G-band flux to compute the flux ratio in equation~\ref{eqn:kepler3_photocentre}.
We apply this model to the Orbital300 set used in section~\ref{sec:results_main} and compare to the infrared-free results obtained there.

\begin{figure}[t]
\begin{center}
\includegraphics[width=0.49\textwidth, angle=0]{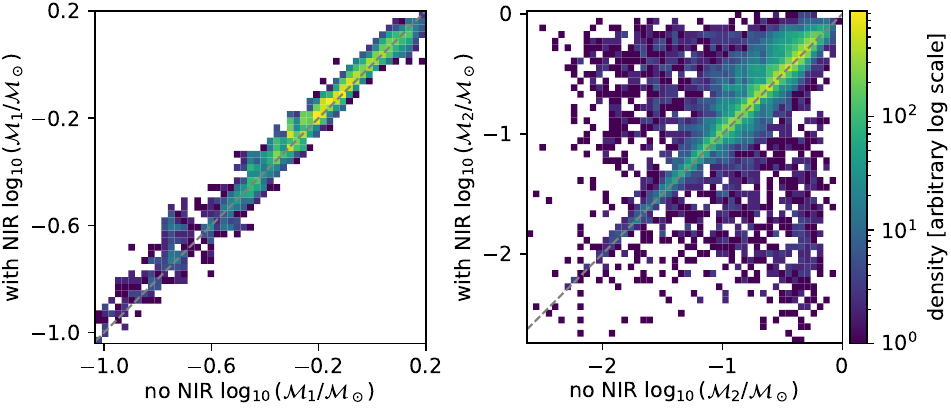}
\caption{Comparison of our mass estimates using infrared photometry (vertical axis) with not using infrared photometry (horizontal axis) on the Orbital300 subset for the primary (left) and secondary (right).
  \label{fig:masses_2D_histogram_nss_3_Orbital_3parsol_withNIR20250923_vs_noNIR20250916}
}
\end{center}
\end{figure}

The overall results are very similar, as we see in figure~\ref{fig:masses_2D_histogram_nss_3_Orbital_3parsol_withNIR20250923_vs_noNIR20250916}. There is, however, a slight tendency for the inclusion of the infrared photometry to increase the estimated masses of both the primaries and the secondaries, by 2.2\% and 3.9\% respectively (median of the mass differences, i.e.\ the bias in equation~\ref{eqn:bias}).
This could be a consequence of neglecting interstellar extinction. If a non-zero extinction is ignored, the system has a brighter intrinsic magnitude than we measure, which generally corresponds to a higher mass. So by ignoring extinction we expect to underestimate the mass. As infrared photometry is less affected by extinction than the visual \gaia\ bands, it is plausible that its inclusion delivers higher mass estimates. It is partly for this reason that we limited our analysis to systems within 300\,\pc. We do expect some extinction in some directions within this horizon, and this interpretation is perhaps confirmed by a slight increase in the difference in the mass estimates with distance, albeit with a weak correlation (Pearson correlation coefficients of 0.01 and 0.10 for \mass1\ and \mass2\ differences respectively).
Yet the above mass increases are small, and
the analysis in section~\ref{results:extinction_correction} also indicates that neglected extinction has only a small effect on the mass estimates. 

Does adding infrared photometry improve the precision of the mass estimates? To address this we
compare the widths of the posteriors using the lower and upper fractional uncertainties. Recall that figure~\ref{fig:nss_3_Orbital_mass_fracunc} shows their distributions without infrared photometry.
Including infrared photometry reduces the fractional uncertainties on \mass1\ by 7\% and 34\% for the lower and upper error bars respectively (median reduction). For \mass2\ they are smaller by $-1$\% for the lower error bars (i.e.\ larger, so slightly less precise) and by 14\% for the upper error bar.

In summary, we find that including infrared photometry gives more precise mass estimates, but doesn't change the estimates by more than a few percent.

\section{Inclusion of spectroscopic orbits}\label{sec:results_noNIR_withSpectro}

\begin{figure}[t]
\begin{center}
  \includegraphics[width=0.49\textwidth, angle=0]{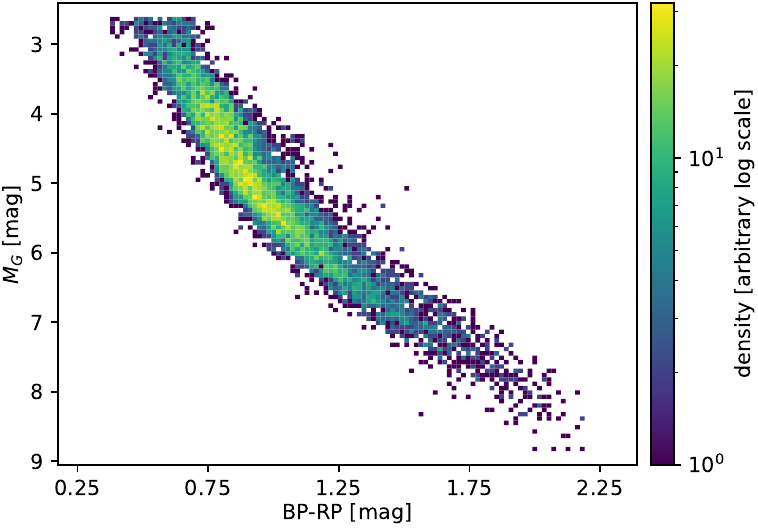}
\caption{Colour--absolute magnitude diagram for the AstroSpectro300 sample assuming zero extinction.
\label{fig:input_data_CQD_nss_4_AstroSpectroSB1+M1_withSpectro_noNIR_20250915}
}
\end{center}
\end{figure}

Some of the stars in the \gdr3\ NSS table have spectroscopic orbital solutions in addition to astrometric orbital solutions. To analyse the impact on the mass inference when using also the spectroscopic measurement (see section~\ref{sec:data_spectroscopic}), we create a new sample single-lined spectroscopic binaries by selecting all those sources lying within 300\,\pc\ that have {\tt nss\_solution\_type} set to {\tt AstroSpectroSB1} and
{\tt combination\_method} flag (section~\ref{sec:comparison}) set to {\tt AstroSpectroSB1+M1}. Retaining only sources within the photometric limits of our forward models (section~\ref{sec:forward_model}), this gives a set of 9674 systems which we call the AstroSpectro300 sample (used already in section~\ref{sec:comparison}). The CAMD of this sample is shown in  figure~\ref{fig:input_data_CQD_nss_4_AstroSpectroSB1+M1_withSpectro_noNIR_20250915}.
Because these sources must have (good) \gaia\ radial velocity spectra, this sample is brighter
than the Orbital300 sample: the 95th percentile in the apparent G magnitude distribution is 12.7\,mag, compared to 15.8\,mag for the Orbital300 sample.

\begin{figure}[t]
\begin{center}
\includegraphics[width=0.49\textwidth, angle=0]{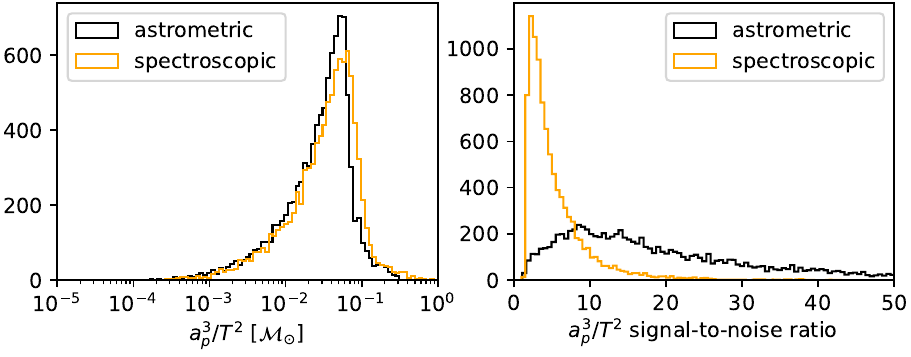}
\caption{Histograms of the distribution of the astrometric and spectroscopic measurements (left) and their \snrs\ (right) for the AstroSpectro300 sample.
\label{fig:input_data_Afunc_Sfunc_nss_4_AstroSpectroSB1+M1_withSpectro_noNIR_20250915}
}
\end{center}
\end{figure}

We now add an extra term for the spectroscopic measurement to the likelihood (equation~\ref{eqn:spectroscopic_measurement}). Everything else is unchanged. Infrared photometry is not used.
The left panel of figure~\ref{fig:input_data_Afunc_Sfunc_nss_4_AstroSpectroSB1+M1_withSpectro_noNIR_20250915} compares the sizes of the spectroscopic and astrometric measurements. The former are larger by 11\% on average (median). This is what we expect, because the non-zero flux contribution from the secondary reduces the astrometric measurement relative to the spectroscopic one (compare equations~\ref{eqn:kepler3_photocentre} and~\ref{eqn:spectroscopic_measurement}),

\begin{figure}[t]
\begin{center}
\includegraphics[width=0.49\textwidth, angle=0]{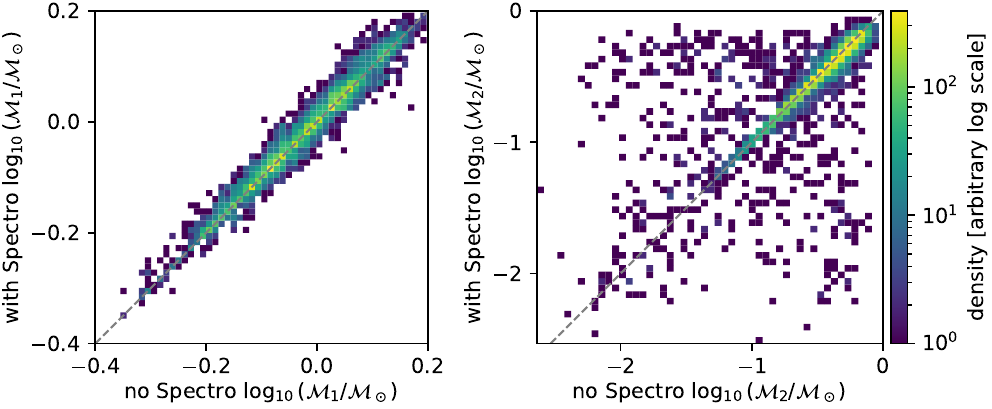}
\caption{Comparison of our mass estimates using spectroscopy (vertical axis) with not using spectroscopy (horizontal axis) on the AstroSpectro300 sample for the primary (left) and secondary (right) shown on a logarithmic density scale.
  \label{fig:masses_2D_histogram_nss_4_AstroSpectroSB1+M1_noNIR_withSpectro20250915_vs_noSpectro20250915}
}
\end{center}
\end{figure}

Figure~\ref{fig:masses_2D_histogram_nss_4_AstroSpectroSB1+M1_noNIR_withSpectro20250915_vs_noSpectro20250915} compares the results of the mass estimates when including spectroscopy with not including it.
Overall we see very little difference. The primary masses are very similar, differing by less than 0.1\% (bias; equation~\ref{eqn:bias}). The scatter (MAR; equation~\ref{eqn:MAR}) is ten times larger, but is still small. The secondary masses are slightly smaller on average (bias) when using spectroscopy, but only by 0.7\%, and the scatter is only 2.7\%.
The fractional uncertainties of these mass estimates are smaller when including spectroscopy, up 40\% on average for the upper fractional uncertainty for the secondaries, but only 2--14\% for the other three fractional uncertainties.

Overall, we find that including the \gdr3\ spectroscopic information for this sample does not significantly change the mass estimates for most systems, and only modestly improves the precision of the estimates. This is an important result, because it implies that \gaia-precision radial velocity follow-up of stars that lack spectroscopic orbits in \gaia\ will not help identify the lowest mass companions.
However, this conclusion only applies to this sample which is quite bright and nearby.
In this sample, the spectroscopic measurement \snr\ is on average four times lower than the astrometric measurement \snr, as we can see in figure~\ref{fig:input_data_Afunc_Sfunc_nss_4_AstroSpectroSB1+M1_withSpectro_noNIR_20250915}, with median values of 16.4 vs 4.1.
More distant stars will generally have poorer astrometric orbits (as the astrometric measurement decreases in size with increasing distance). The \gaia\ epoch radial velocities are not very precise -- which leads to the low \snr\ in the spectroscopic measurement -- so higher precision radial velocities would help in follow-up.

\section{Limitations of the method}\label{sec:discussion}

To determine intrinsic uncertainties in our inference method, we used it to estimate the masses of simulated binary systems. This is described in appendix~\ref{sec:performance_simulated_binaries}. The systems and the \snr\ of the data are broadly similar to the Orbital300 sample.
We found that the primary masses can be estimated with an accuracy (MAR, equation~\ref{eqn:MAR}) of 3\% and a bias less than $-0.2$\% (that is, a near negligible underestimation). For the secondary mass the MAR is 5\% and the bias is $-3$\%. This is quite a lot better than what we achieved on the Orbital300 sample.

There are a number of reasons why the performance on real data is probably worse that what these simulations yield. These could include:
\begin{enumerate}
\item mismatch between the PARSEC models and real stars; 
\item assumption of zero interstellar extinction; 
\item non-Gaussian likelihoods and/or incorrectly-estimated uncertainties.
\end{enumerate}

The first of these issues is a known problem and could well be the dominant issue.
In particular, our extrapolation of the fit below the lower mass limit of the PARSEC models should ideally be replaced by a brown dwarf and exoplanet model.

The second issue is a problem in principle, but our analyses using the GSP-Phot extinction values (masses systematically overestimated by less than 2\%; see section~\ref{results:extinction_correction}), as well as the results from including infrared photometry (see section~\ref{sec:results_withNIR_noSpectro}), suggest neglected extinction has little impact for our nearby sample (within 300\,\pc). We chose not to adjust the photometry using one of the various published 3D dust maps, because these tend not to fit very well in the solar neighbourhood, plus they are not designed to give accurate results for individual stars.

Concerning the third issue, we found that inflating the \gaia\ photometric uncertainties by a factor of five changes the inferred masses by much less than 1\% on average, and changes the \snr\ in these masses by less than 1\% on average.
On the other hand, non-Gaussianity in the sometimes noisy astrometric measurements may be an issue (figure~\ref{fig:nss_3_Orbital_input_data}), not least because it is computed from the ratio of just two quantities (equation~\ref{eqn:likelihood_Afunc_variance}), so central limit theorem convergence to a Gaussian may not apply. This uncertainty estimate has also neglected the contribution from the scatter in the forward models, which could be a limiting factor when the astrometric measurements are more precise..

Some studies, such as \cite{2024A&A...688A..44W} and \cite{2025MNRAS.536.2485W}, 
adopt a simplified version of the method in this paper. By assuming $\flux2 \ll \flux1$, the observed photometry allows the mass of the primary to be determined independently of the secondary, and the simplified version of equation~\ref{eqn:kepler3_photocentre} then yields the mass of the secondary. However, this assumes we already know that the secondary is faint. If the single-star main sequence were an arbitrarily thin locus, the absolute photometry would betray the presence of a bright secondary. But due to age and metallicity variations the main sequence is not arbitrarily thin, and if the photometry is not dominated by the primary then the single star inference from the photometry will give an incorrect primary mass. Our method avoids this assumption to produce a self-consistent solution for the masses and fluxes of both components.

\section{Summary and conclusions}\label{sec:conclusions}

We have developed a method to infer the masses of both components of an unresolved binary using the photocentre semi-major axis, orbital period, parallax, and photometric flux measurements.
This involves using a forward model to map the relationship between the fluxes and the two masses and two latent variables (system age and metallicity). We fit this model using PARSEC stellar models.
We sample the resulting four-dimensional posterior using a Monte Carlo method.

We tested our method on \gaia\ astrometry and photometry (three bands) for binaries listed in the NSS table in \gdr{3}. Our main sample (Orbital300) is a set of 20\,334 systems within 300\,\pc\ that have system photometry consistent with the stellar main sequence. We can estimate primary masses with median precisions (1-$\sigma$-like uncertainties) of 6\% and secondary masses with median precisions of 25\%. For a significant fraction of the systems, however, the posterior for the secondaries extends to zero with non-negligible probability density, indicating a large lower uncertainty on the secondary mass.

The numbers above do not refer to deviations from the true masses. 
Dynamical-only `true' mass estimates are rarely available, because they require additional information available only for a limited class of object (in particular, double-lined spectroscopic binaries).
We could, however, compare our results to mass estimates obtained by \cite{2023A&A...674A..34G}.  For the Orbital300 sample, the primary masses are quite similar: the scatter (median absolute residual in the log masses) is 6.4\%, which includes a bias of $-5.5$\% (our estimates are lower). The secondary masses show larger deviations: a scatter of 39\%, which includes a bias of $-38$\%. However, on a sample of brighter stars (AstroSpectro300), the agreement is much better. For the secondaries the scatter is 14\%, which includes a bias of just $+1.6$\%. For this, \cite{2023A&A...674A..34G} also used the spectroscopic orbits, whereas we did not.

Our method accommodates the initially unknown flux contribution of the secondary that affects the size of the observed semi-major axis. We found that neglecting this underestimates the mass of the secondary
by 9\% on average (median),
but by 40\% or more for 10\% of the sources.
Neglecting this could therefore be a source of false positive low mass companions.

We did not account for interstellar extinction, and this could in principle lead to mass underestimates.
However, we found that correcting for the extinction using values from \gdr{3} had very little impact on our mass estimates. This is supported by the fact that including infrared photometry  (J, K$_s$, W1, W2) in the inference does not change the mass estimates by more than a few percent. This additional photometry does, however, produce more precise mass estimates (narrower posteriors).

Including \gaia's spectroscopic orbit in the inference hardly changes the mass estimates. This suggests that \gaia-like spectroscopic follow-up of these systems will not improve the mass estimates. More precise radial velocities at the appropriate epochs will help, however.

Using our method on the larger and more complete set of astrometry that will be published in \gdr{4}, we plan to estimate the masses of both components of many more binaries using only the astrometric orbit and photometric data.
With this we hope in particular to identify substellar companions without being significantly affected by contamination of stellar companions having the same astrometric signal.

\begin{acknowledgements}

This work uses data from the European Space Agency (ESA) mission
\gaia\ (\url{https://www.cosmos.esa.int/gaia}), processed by the \gaia\
Data Processing and Analysis Consortium (DPAC,
\url{https://www.cosmos.esa.int/web/gaia/dpac/consortium}). Funding for the DPAC
has been provided by national institutions, in particular the institutions
participating in the \gaia\ Multilateral Agreement.
This publication also uses data products from the Two Micron All Sky Survey (2MASS), a joint project of the University of Massachusetts and the Infrared Processing and Analysis Center/California Institute of Technology, funded by the National Aeronautics and Space Administration and the National Science Foundation, and from
the Wide-field Infrared Survey Explorer (WISE), a joint project of the University of California, Los Angeles, and the Jet Propulsion Laboratory/California Institute of Technology, funded by the National Aeronautics and Space Administration.
We also made use of
CDS's Simbad \citep{2000AAS..143....9W}, NASA's \href{http://adsabs.harvard.edu/abstract_service.html}{ADS},
and matplotlib \citep{Hunter:2007} in this work.

\end{acknowledgements}

\bibliographystyle{aa}
\bibliography{bibliography}

\appendix

\section{Performance on simulated astrometric binaries}\label{sec:performance_simulated_binaries}

To determine the limiting performance of our inference method, we use it here to infer the masses of simulated astrometric binaries. 
To generate our sample, we first draw parameters of individual stars at random from uniform distributions in $\log_{10}\mass{}$, $\log_{10}(\age/\yr)$, and \met. The limits of these distributions are the range of the forward model grids (defined at the beginning of section~\ref{sec:forward_model}), except for the lower mass limit, which is set to a lower mass of 0.01\,\msun.
We then use the fitted forward models described in section~\ref{sec:forward_model} to predict the photometry for each star, including the extrapolation for the lower mass stars.

We generate binaries by randomly pairing stars.
This is done by drawing a period at random for each system from a distribution uniform in log period between 100 and 1500 days, then computing the SMA from equations~\ref{eqn:kepler3} and~\ref{eqn:sma_photocentre} and the 
true (noise-free) astrometric measurement from equation~\ref{eqn:kepler3_photocentre}.
We assign a parallax to each system at random 
by drawing a distance from a uniform space density distribution ($P(r) \propto r^2$) up to 1\,kpc with a 
lower limit set for each system to ensure that the photocentre SMA is below 100\,mas and so would be unresolved by \gaia.

To generate noisy data for the inference we add Gaussian noise to all inputs. For simplicity we assume fixed \snrs\ for each input with the following values: 9.9 for the astrometric measurement (this is dominated by the uncertainty in the SMA);
200 for the G band flux; 100 for each of the BP and RP band fluxes; 150 for the parallax.
This \snr\ in the astrometric measurement is close to the median of the Orbital300 sample of 9.4.
The high, constant \snrs\ for the \gaia\ fluxes essentially assume that the uncertainties are dominated by systematic errors rather than photon noise, which is consistent with the NSS sources being bright.
The uncertainties corresponding to these \snrs\ are also used as the standard deviations $\sigma_A$ (equation~\ref{eqn:likelihood_Afunc_variance}) and $\sigma_{f_X}$ (equation~\ref{eqn:likelihood_flux_variance}) in the likelihoods.

\begin{figure}[t]
\begin{center}
\includegraphics[width=0.49\textwidth, angle=0]{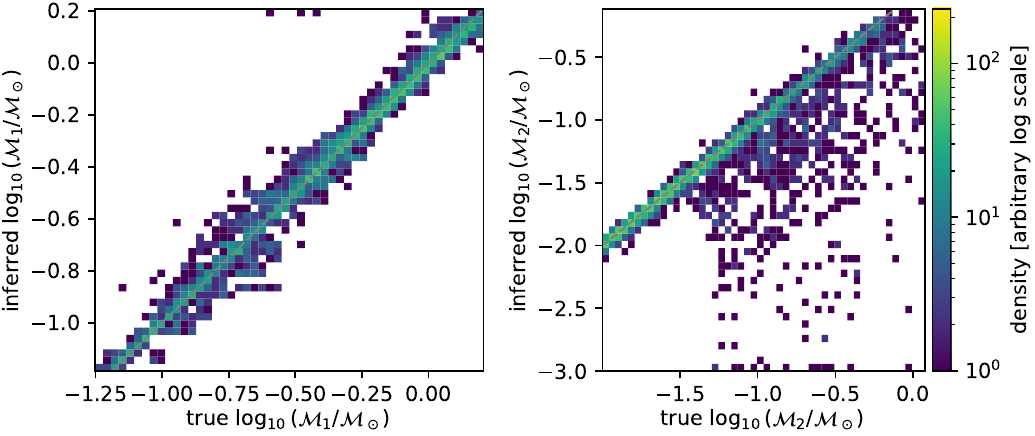}
\caption{Performance of mass estimation on the simulated data set using astrometry and \gaia\ photometry.
  Note that in each panel, the horizontal and vertical axis do not span a common range.
  \label{fig:mass_comparison_simulated}
}
\end{center}
\end{figure}

We compute and summarize the posteriors as done for the real data (section~\ref{sec:posterior_pdf_sampling}). The median of the posteriors for the primary and secondary masses are compared to the true masses in figure~\ref{fig:mass_comparison_simulated}.  For the primaries, the bias and MAR $\log_{10}\mass1$ are $-0.0008$\,dex and $0.013$\,dex respectively, that is, negligible bias and a median scatter of 3\%. For the secondaries the bias and MAR are $-0.0136$\,dex and $0.0236$\,dex respectively. The former means the inference underestimates the masses by 2\% on average, it being generally lower for low mass systems and higher for higher mass systems.

If we include the four infrared photometric bands as additional measurements, the results are very similar. The biases for the primaries are secondaries are quite similar, as is the MAR on the secondaries. But the MAR on the primaries is three times smaller when using infrared photometry. The \snrs\ on the mass estimates of the primaries are also about twice as large, for both the lower and upper uncertainties, when using the infrared photometry.

This performance on simulated data is considerably better than the consistency we see in section~\ref{sec:comparison} between our real results and other literature estimates, in particular for the secondaries. Possible reasons for this are discused in section~\ref{sec:discussion}.
The simulations obviously present an ideal situation, in which the measurements -- one astrometric measurement and the three fluxes -- differ from the true values by only random Gaussian noise.
This nonetheless indicates that in principle our method of determining the four system parameters -- two masses, age, and metallicity -- from the four inputs can infer accurate masses, even for the secondary.
These results on simulated data represent an approximate limit on the mass estimation accuracy that we can achieve on real data.

\section{Contamination of stellar companions in detecting exoplanet companions}\label{sec:contamination}

A follow-up goal of this work when \gdr{4} data become available is to estimate the masses of exoplanet companions in unresolved astrometric star--planet binaries. As the planet contributes negligible light to the integrated photometry, the estimation of the mass of both the stellar and planet mass from the astrometric measurement and the flux is easier than considered in this paper, because we can then assume $f_1/f_2 = \infty$ in equation~\ref{eqn:kepler3_photocentre} and need the forward model to map mass to luminosity for the stellar primary only.

However, we do not know a priori that the secondary is a planet.  As discussed in section~\ref{sec:breaking_degeneracy}, a near equal-mass and equal-luminosity star--star binary can have an arbitrarily small photometric SMA and thus be confused as a star--planet binary. Although the method outlined in this paper is designed to overcome this, 
the mass uncertainties on the secondary can nonetheless be large (figure~\ref{fig:nss_3_Orbital_mass_fracunc}).
But how often do `contaminating' star--star binaries occur in practice?

Here we make some simple assumptions about the expected populations of star--star and star--planet binaries to investigate statistically how likely it is that the former have photometric SMAs similar to the latter. We look at stars with masses between 0.8 and 1.5\,\msun, which corresponds roughly to F and G dwarfs.

\begin{figure}[t]
\begin{center}
  \includegraphics[width=0.49\textwidth, angle=0]{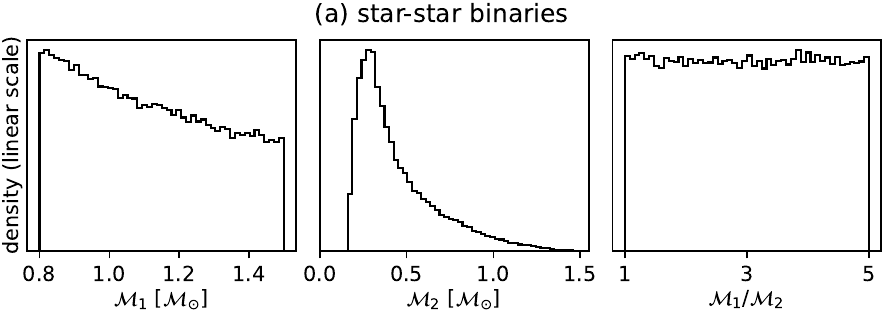}
  \includegraphics[width=0.49\textwidth, angle=0]{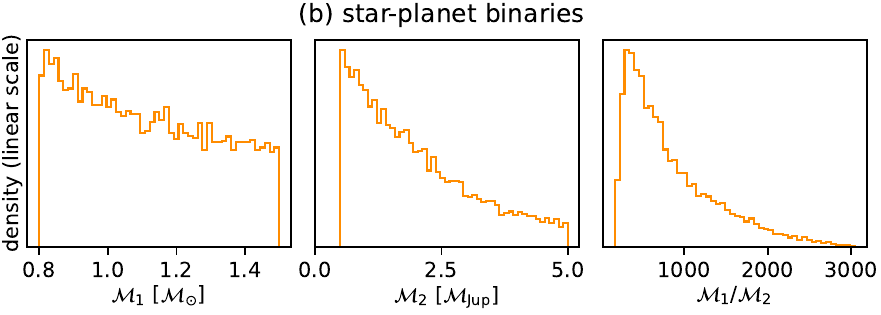}
\caption{Distribution of the masses and mass ratio in the simulated star--star systems (top), and the simulated star--planet systems (bottom). 
  \label{fig:syssims_mass_distributions}
}
\end{center}
\end{figure}

To model both types of binaries, we draw the mass of the primary from a truncated power law
$P(\mass1) \propto \mass1^{-1}$. For the star--star binaries we then draw the mass ratio from a flat distribution and use this to assign the mass of the secondary. The resulting distributions are shown in the upper panel of figure~\ref{fig:syssims_mass_distributions}. For the star--planet binaries we draw the planet mass from a gamma distribution truncated between 0.5 and 5.0\,\mjup. The resulting distributions are shown in the lower panel of figure~\ref{fig:syssims_mass_distributions}
We simulate ten times as many star--star as star--planet binaries to reflect a higher occurence of stellar companions over massive planet companions around F and G stars \citep{2023AJ....166...11P}.

\begin{figure}[t]
\begin{center}
  \includegraphics[width=0.49\textwidth, angle=0]{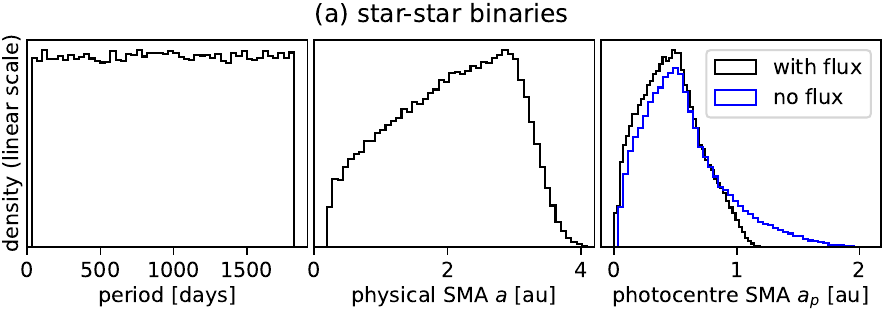}
  \includegraphics[width=0.49\textwidth, angle=0]{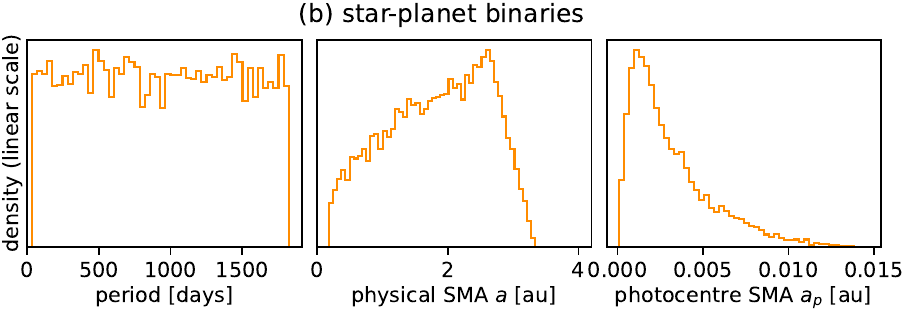}
\caption{Distribution of the period and SMAs in the simulated star--star systems (top), and the simulated star--planet systems (bottom).The two left panels span a common range, as do the two centre ones, but the right ones do not. The blue curve in the upper right panel shows the distribution if we erroneously neglect the flux from the secondary in computing the photocentre of the SMA.
  \label{fig:syssims_period_SMA_distributions}
}
\end{center}
\end{figure}

The orbital period for each system is drawn from a uniform distribution between 30 days and 5 years.
From the period and masses we compute the physical SMA from Kepler's third law (equation~\ref{eqn:kepler3}).
To compute the SMA of the photocentre ($\sma{p}$) for the star--star binaries, we need to take into account the flux contributions from both components. To do this we first assign
a log age and a metallicity for the system by drawing these quantities from uniform distributions with limits equal to those of the forward model grid (section~\ref{sec:forward_model}). We then use the forward models to predict the flux for both components and finally equation~\ref{eqn:sma_photocentre} to compute $\sma{p}$. The resulting distributions are shown in the upper panel of figure~\ref{fig:syssims_period_SMA_distributions}.

For the star--planet binaries we assume the planet contributes no flux, so we don't need to simulate its photometry. The SMA of the photocentre is given by equation~\ref{eqn:kepler3_photocentre} with $f_1/f_2 = \infty$. The period and SMA distributions for these binaries are shown in the lower panel of figure~\ref{fig:syssims_period_SMA_distributions}.

\begin{figure}[t]
\begin{center}
\includegraphics[width=0.49\textwidth, angle=0]{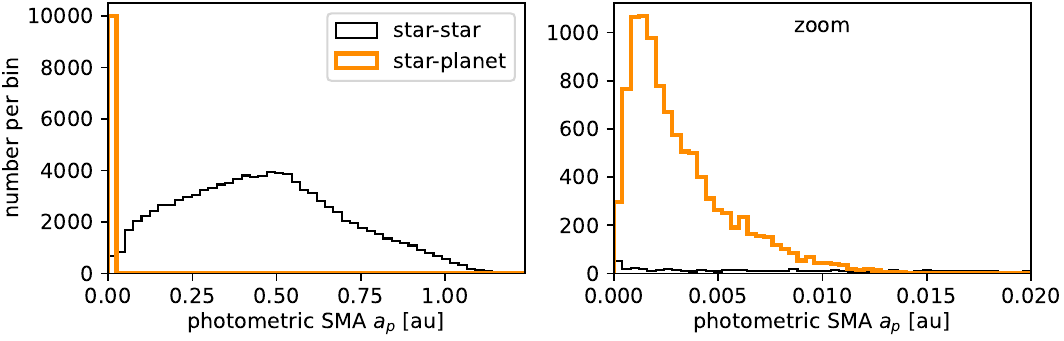}
\includegraphics[width=0.49\textwidth, angle=0]{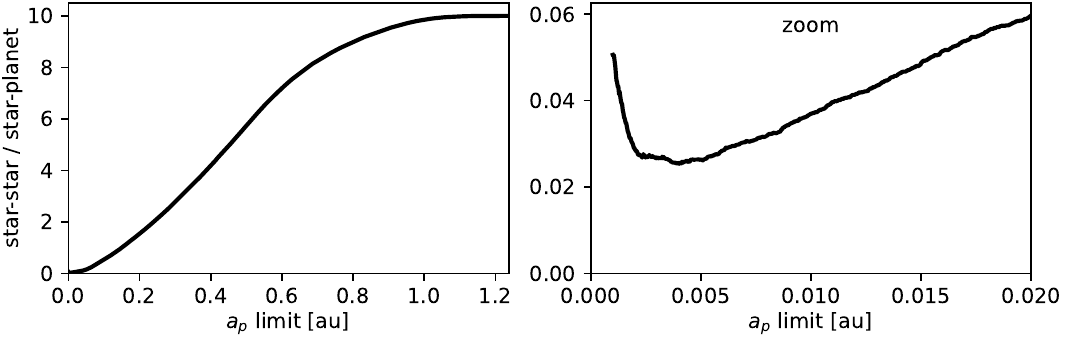}
\caption{Comparison of the photometric SMA distributions for the simulated star--star and star--planet systems.
The right panel is a zoom of the left panel in both rows.
The top row shows the actual number of systems per bin (taking into account that star--planet binaries are ten times rarer). The bottom row shows the ratio of the number of star--star to star--planet systems that have a photometric SMA less than the value shown on the horizontal axis.
  \label{fig:syssims_SMA_distributions_compared}
}
\end{center}
\end{figure}

It is already apparent from the above figures that the photometric SMAs of the star--planet binaries are usually much smaller than those of the star--star binaries. This is more obvious in figure~\ref{fig:syssims_SMA_distributions_compared}, which plots them together. The bottom row shows the ratio of the number of systems of each type with photometric SMA below some limit. We see that below 0.01\,\au, there are only 0.037 star--star binaries for every star--planet binary.
So although is possible for a near-equal mass star--star binary to have a very small photometric SMA, it is rare that it is small enough to mimic a star--planet binary.
This result is quite robust to changes both in the shapes of the mass distributions adopted and to the lower period limit.
Even if we now assume stars--star binaries outnumber star--planet ones by a factor of 50, and so scale up the star--star curve by a factor of five, the above fraction becomse 0.18, a `contamination' rate of 1 in 6.5.

A more detailed analysis of the expected stellar binary contamination in the \gdr{4} exoplanet yield is given by \cite{2025arXiv251104673L}, and is also studied in \cite{2023AJ....165..266M}.

\end{document}